\documentclass[iop,twocolumn]{emulateapj}

\usepackage{psfig}
\usepackage{color}
\usepackage{natbib}
\usepackage{ulem}
\usepackage{soul}

\def\a0208{PKS~0208-512}
\def\g0235{PKS~0235+164}
\def\ba0402{PKS~0402-362}
\def\ca0454{PKS~0454-234}
\def\m0528{PKS~0528+134}
\def\oj287{OJ~287}
\def\cf273{3C~273}
\def\cb279{3C~279}
\def\df1510{PKS~1510-089}
\def\e2052{PKS~2052-474}
\def\f2142{PKS~2142-75}
\def\h2155{PKS~2155-304}
\def\thesource{{3C~454.3}}
\def\j0531{PMN~J0531-4827}
\def\k23455{PKS~2345-1555}

\newcommand{\ang}{\mbox{$\:$\AA}}

\def \deg{^\circ}

\def \pow10#1{\times 10^{#1}}

\def\mgii{{Mg~{\sc II}}}
\def\siii{{Si~{\sc III]}}}
\def\ciii{\ifmmode {\rm C{\sc III]}} \else C~{\sc III]}\fi}
\def\ha{\ifmmode {\rm H}\alpha \else H$\alpha$\fi}
\def\civ{\ifmmode {\rm C{\sc IV}} \else C~{\sc IV}\fi}
\def\hg{\ifmmode {\rm H}\gamma \else H$\gamma$\fi}
\def\hb{\ifmmode {\rm H}\beta \else H$\beta$\fi}
\def\gsim{\lower 2pt \hbox{$\, \b~uildrel {\scriptstyle >}\over
{\scriptstyle \sim}\,$}}
\def\lsim{\lower 2pt \hbox{$\, \b~uildrel {\scriptstyle <}\over
{\scriptstyle \sim}\,$}}
\def\smarts{\textit{SMARTS}}
\def\fermi{\textit{Fermi}}

\def\fflux{erg s$^{-1}$ cm$^{-2}$}
\def\pflux{ph s$^{-1}$ cm$^{-2}$}

\def\g{$\gamma$}
\def\b{$\beta$}

\slugcomment{Accepted to the Astrophysical Journal}

\shortauthors{Isler et al.}
\shorttitle{SaMOSA: Emission Line Variability}

\begin{document}

\title{The SMARTS Multi-Epoch Optical Spectroscopy Atlas (SaMOSA): \\ Using Emission Line Variability to Probe the Location of the Blazar \\Gamma-emitting Region}

\author{Jedidah C. Isler\altaffilmark{1,2}, C.M. Urry\altaffilmark{2,3},  C. Bailyn\altaffilmark{2,4}, P. S. Smith\altaffilmark{5}, P. Coppi\altaffilmark{2},
M. Brady\altaffilmark{2},\\
E. MacPherson\altaffilmark{2},
I. Hasan\altaffilmark{2},
M. Buxton\altaffilmark{2}
}

\altaffiltext{1}{Chancellor's Faculty Fellow, Syracuse University, Department of Physics, Syracuse, NY 13214;jcisler@syr.edu}
\altaffiltext{2}{Department of Astronomy, Yale University, PO Box 208101, New Haven, CT 06520-8101}
\altaffiltext{3}{Department of Physics and Yale Center for Astronomy and Astrophysics, Yale University, PO Box 208121, New Haven, CT 06520-8120}
\altaffiltext{4}{Yale-NUS College, 6 College Avenue, East, Singapore, 138614}
\altaffiltext{5}{Steward Observatory, University of Arizona, 933 N. Cherry Avenue, Tuscon, AZ 85721}


\begin{abstract}
We present multi-epoch optical spectroscopy of seven southern \fermi-monitored blazars from 2008 - 2013 using the Small and Medium Aperture Research Telescope System (\smarts), with supplemental spectroscopy and polarization data from the Steward Observatory. We find that the emission lines are much less variable than the continuum; 4 of 7 blazars had no detectable emission line variability over the 5 years. This is consistent with photoionization primarily by an accretion disk, allowing us to use the lines as a probe of disk activity. Comparing optical emission line flux with \fermi\ \g-ray flux and optical polarized flux, we investigate whether relativistic jet variability is related to the accretion flow. In general, we see no such dependence, suggesting the jet variability is likely caused by internal processes like turbulence or shock acceleration rather than a variable accretion rate. However, three sources showed statistically significant emission line flares in close temporal proximity to very large \fermi\ \g-ray flares. While we do not have sufficient emission line data to quantitatively assess their correlation with the \g-ray flux, it appears that in some cases, the jet might provide additional photoionizing flux to the broad line region, which implies some \g-rays are produced within the broad line region, at least for these large flares.
\end{abstract}

\section{Introduction}
Blazars are radio-loud Active Galactic Nuclei (AGN) whose relativistic jet are pointed at small angles with respect to our line of sight \citep{antonucci93, Urry95}. This orientation makes blazars an ideal laboratory for the study of jet physics, due to Doppler beaming that greatly increases the source brightness and decreases the variability timescale.

Blazars are observable across the electromagnetic spectrum. Their broad band spectral energy distribution (SED) has two characteristic peaks: one at low frequencies (infrared -- soft X-ray) due to synchrotron radiation and one at high frequencies (MeV -- TeV), likely due to inverse Compton scattering.

Since the launch of \fermi\ in 2008, the GeV behavior of blazars has been studied in unprecedented detail. Time variability studies have constrained emission models by identifying correlations between the two spectral peaks, using high cadence observations with \fermi\ Large Area Telescope (LAT) and coordinated multiwavelength campaigns from radio to TeV \citep[e.g.][]{Bonning09, Abdo09, DAmmando09, Ghisellini09, Ackermann10, Pacciani10, Poutanen10, Fermi1LAC10, Orienti11, Marscher11, Aller11, Jorstad12, Sbarrato12, Bonning12,  Agudo12, Sandrinelli13, Chatterjee13, Nalewajko13, Hess13, Ghisellini13, Tavecchio13, Cao13}.

However, the thermal components of AGN, namely the accretion disk and broad line region, are still energetically relevant.  For example, the disk could, via magnetic interactions, contribute to the initial collimation of the relativistic jet \citep{BlandfordPayne82}, although the exact mechanism is not well understood. Furthermore, in some blazars, the accretion disk has been shown to contribute a significant fraction of the radiation energy density on sub-pc scales \citep[e.g.][]{Ghisellini09}. 

While many studies of the total flux variability of blazars have been undertaken, similar high-quality, multi-epoch spectroscopic studies have been more challenging. Thus, in this work we have focused our analysis on the emission line variability, as the emission lines could be an appropriate proxy for the disk emission, which is often swamped by jet continuum in high flaring states. We have another manuscript in preparation that will address the continuum variability of these blazars and what, if any, relationship can be drawn to the emission line variability properties discussed here (Isler et al. 2015, \textit{in prep}). 

Early spectroscopic studies of blazars showed emission line variability on month to year timescales \citep[e.g.,][]{ulrich93, Falomo94, Koratkar98}, but were not carried out in conjunction with \g-ray observations, so the impact of the jet on these sources could not be easily investigated. In principle, the relativistic jet could contribute additional photoionizing flux to the emission lines, causing significant jet-cooling within the broad line region, provided the photoionizing emission arises on smaller spatial scales than the broad line gas. This geometry is required by the forward beaming of the jet emission, very little of which is directed backwards. We are now able to compare directly (and simultaneously) the multi-epoch optical spectroscopic observations of the broad line region flux to the jet flux using \fermi, \smarts\ and Steward Observatory data. If a relationship is found, the spatial scale of the \g-emitting region can be tightly constrained.

Simultaneous emission line variability studies in \fermi-monitored blazars have generated mixed results. Among a set of similar \g-ray and optically bright, variable quasars, no emission line variability was detected in PKS~1222+216 or 4C 38.51 \citep{Smith11, Farina12, Raiteri12}. By contrast, \thesource\ had factors of two emission line variation roughly coincident with high \g-ray emission levels \citep{Isler13, LT13}. These studies underscore the importance of monitoring blazar emission line variability, but represent too limited a sample of the total population to draw statistical conclusions about blazars as a class.

We measure the emission line behavior of 7 blazars to investigate the presence of short timescale (weeks to months) emission line variability, and assess if that variability is temporally related to \fermi\ \g-ray flares. This study could provide a direct test of the contribution of photoionizing flux from the jet to the broad line region, and if a correlation is found, observationally constrain the location of the \g-emitting region to be within the broad line region for those flares.

In Section~\ref{sec:sammeth}, we discuss the sample selection, observational program and data analysis. In Section~\ref{sec:samres}, we present the \fermi\ \g-ray and emission line flux light curves and we define empirical line flares as well as the measures for statistical variability. In Section~\ref{sec:samcols} we analyze the optical linear polarization at the time of observation as an additional measure of the non-thermal jet contribution to the optical waveband. We discuss the emission line behavior of the sample and its implications for current jet dissipation models in Section~\ref{sec:samdisc}; main conclusions are summarized in Section~\ref{sec:samcon}. The following cosmological parameters were used throughout this work: H$_0$~=~71 km s$^{-1}$ Mpc$^{-1}$, $\Omega_m$~=~0.27, $\Lambda_0$~=~0.73, and  q$_0$ = -0.6.

\section{Sample Selection and Data Analysis}\label{sec:sammeth}
Since 2008, we have carried out optical spectroscopic monitoring of approximately 30 \fermi\ \g-ray bright blazars at the queue-scheduled Small and Medium Aperture Research Telescope System (\smarts) in Cerro Tololo, Chile.  

\subsection{Sample Selection}
The \smarts\ Multi-epoch Optical Spectroscopy Atlas (SaMOSA) was based on the original \fermi-LAT `bright source list' released just before launch in 2008, including those blazars with declination $<$ 20$\deg$, given the location of the \smarts\ telescopes. 
The original list included Flat Spectrum Radio Quasars (FSRQs) and BL Lac objects (BLLs), which are distinguished by whether broad emission lines are present at levels greater or less than 5\ang, respectively \citep{Angel80}. In subsequent years, the SaMOSA list was expanded to include newly-flaring \fermi\ blazars, defined as having F$_\gamma$ (E~$>$~100~MeV) $\ge$ 1$\pow10{-6}$ \pflux, and flaring FSRQs publicized on the Astronomers Telegram\footnotemark[1]. We did not include BLLs in the final analysis because no emission lines were detected in their optical spectra. Since the purpose of the current study is to understand broad line variability, we only include objects for which at least five epochs of spectroscopy are available 
\footnotetext[1]{http://www.astronomerstelegram.org} for a total of 7 FSRQs. Table~\ref{tab:samshort} lists the SaMOSA sample, \fermi\ identifier, redshift, number of observations and emission lines included in the analysis.

\subsection{The \smarts\ Optical/Near-infrared Photometry and Optical Spectroscopy}
The \smarts\ optical/near-infrared (OIR) photometry is obtained nightly with the 1.3m + ANDICAM, a dual-channel imager with a dichroic that simultaneously feeds an optical CCD and infrared IR imager, with spectral coverage from 0.4 - 2.2 $\mu$m. Data analysis for \smarts\ OIR photometry is described in \citet{Bonning12}, so we briefly summarize here. \smarts\ differential photometry is obtained by using optical comparison stars calibrated to Landolt standards on photometric nights. The reported \smarts\ infrared magnitudes are calibrated using Two Micron All Sky Survey (2MASS) magnitudes \citep{Skrutskie06} of at least one secondary star in the same field as the blazar. In both optical and infrared bands, the photometric uncertainties are dominated by the random errors in the comparison star magnitude, and the 1$\sigma$ uncertainty is reported. In the infrared, the dominant uncertainty is the calibration to the 2MASS magnitude. OIR finding charts and comparison star magnitudes for \ba0402, \ca0454, \f2142\ and \e2052\ are provided in the Appendix; \smarts\ finding charts for the remaining sources can be found in \citet{Bonning12}. OIR photometry and finding charts for all \smarts-monitored sources are publicly available via our website\footnotemark[2].
\footnotetext[2]{http://www.astro.yale.edu/smarts/glast/home.php}

\smarts\ optical spectroscopy was obtained with the 1.5m + Cassegrain spectrograph (RCSpec) at an \textit{f}/7.5 focus with plate scale 18.$^{\prime\prime}$1 mm$^{-1}$ and a LORAL 1K (1200 $\times$ 800) CCD. The primary grating for this study has first order resolution of 17.2~\ang, with spectral coverage of 6600~\ang\ and 2$^{\prime\prime}$ slit width. In the optimal case, spectroscopic data were obtained approximately bi-weekly, depending on weather conditions and source visibility in Cerro Tololo, although in many cases significantly less data was acquired. No second order corrections were applied to the spectra, as previous analysis yielded a contamination rate of 8\% or less \citep{Isler13}, which is insignificant compared to the errors introduced by the flux calibration. Nevertheless, the systematic error introduced by the contaminating continuum light is difficult to model, especially in the case when the variability amplitude is larger in the blue than the red part of the continuum. In any case, this contamination applies only to emission lines with observed wavelength $\gtrsim$ 7000\ang, so that H$\alpha$ in \df1510\ is the only emission line affected by this contamination. 

The optical spectroscopic data reduction here is similar to that described by \citet{Isler13}. We use the \textit{MPFIT} package \citep{Markwardt09} to measure emission line equivalent width by fitting a Gaussian to the emission line above the continuum and minimizing the $\chi^2$ statistic. A linear fit to the continuum was applied on each side of the line over 100 \ang. We calculated the noise per pixel by combining the uncertainties from bias subtraction, sky correction, and aperture extraction. 

The uncertainty in the equivalent width was estimated by running 500 Monte Carlo simulations of each fitted line, including the measured noise in the count rate of each pixel (as described above). For each Monte Carlo simulation, the emission line was fitted and the values of the equivalent width were calculated. The reported error of the equivalent width of each line is the standard error on the 500 equivalent width simulations. 

Emission line flux was derived from the equivalent width and  \smarts\ B-, V- or R-band photometry, depending on the observed location of the line in the spectrum. Emission lines with rest-frame line center, $\lambda_{obs}$ $<$ 5000~\ang\ were calibrated with the B-band ($\lambda_{eff}$~=~4400~\ang), line centers 4999 $<$  $\lambda_{obs}$ $<$ 5999~\ang\ were calibrated with the V-band ($\lambda_{eff}$~=~5500~\ang), and $\lambda_{obs}$ $>$ 6000~\ang\ were calibrated with the R-band ($\lambda_{eff}$~=~6600~\ang). No observed line was more than 500~\ang\ from the relevant effective wavelength in any case. Table~\ref{tab:samobs} lists the targets, observations, emission line equivalent widths with their associated uncertainties, as well as B-, V-, R- and J-band magnitude with associated 1$\sigma$ uncertainty for the SaMOSA sample.

\subsection{Steward Observatory Optical Spectroscopy and Polarimetry}
Since the launch of Fermi, the Steward Observatory of the University of
Arizona has carried out regular, publicly available optical
spectrophotometry and linear spectropolarimetry of a large sample of
\g-ray-bright blazars using the Bok 2.3m and Kuiper 1.54m telescopes
\citep{Smith09}.  The SPOL spectropolarimeter \citep{Schmidt92} is
used for this monitoring program.  Observations are made in first order
using a 600 l mm$^{-1}$ grating, yielding a spectral range of 4000~-~7550 \ang\ and
resolution of 15-20 \ang.

Sky-subtracted spectra and broad-band (5000-7000 \ang) polarization
measurements derived from the spectropolarimetry from 2008-2013 were
obtained from the publicly accessible Steward Observatory website\footnotemark[3].  \footnotetext[3]{http:/www.james.as.arizona.edu/$\sim$psmith/Fermi/} The
optical spectroscopy was then reduced in a similar fashion as the SMARTS
spectroscopy. The noise per pixel array for the Steward spectroscopy was estimated \textit{a posteriori} using the gain and read noise. This array was then scaled by a factor of 10 to approximate the root-mean-square in the observed spectrum. The scaling of the uncertainties provides a very conservative estimate of the error in the lines, and likely overestimates the noise in the spectrum. The broad-band polarization measurements were used to derive V-band
polarized flux densities by multiplying the fractional linear polarization
(P) by the V-band flux density (F$_V$).

There is a small systematic offset, of order(0.1) dex, in the line fluxes reported by Steward and \smarts, likely due to differences in the comparison stars used for the analysis. The (logarithmic) offsets from the \smarts\ data are applied to the light curves of \ba0402\ (\mgii: -0.1203), \ca0454\ (\mgii: -0.2493), and \df1510\ (H\g: +0.0481, H\b: -0.0501); the other four sources included in the SAMOSA sample contain only \smarts\ data so no cross-calibration is necessary.

\subsection{\fermi\ \g-ray Fluxes}
\fermi/LAT data were obtained from the first \smarts\ photometric observation for each source through 2013 July 01 (MJD 56474), via the \fermi\ Science Support Center website\footnotemark[4].
Pass 7 reprocessed data (event class 3) were analyzed using \fermi\ Science Tools (v9r33p0) with scripts that automate the likelihood analysis.
Galactic diffuse models (gll\_iem\_v05\_rev1), isotropic diffuse background (iso\_p7v6source) and instrument response functions (P7REP\_CLEAN\_V15) were utilized in the analysis. 
Data were constrained to time periods where the zenith angle was less than 100$^\circ$ to avoid Earth limb contamination, and photons to within a 10$^\circ$ region centered on the source of interest.

The \fermi\ \g-ray spectra of each object were modeled as a power law or log-parabola, according to spectral type listed in the 2FGL catalog, with the photon flux and spectral index as free parameters. 
 \fermi\ light curves were integrated in one-day bins to match the average \smarts\ photometric cadence, and an integral \fermi\ \g-ray flux above 100MeV, F$_\gamma$ is reported. \fermi\ fluxes for which TS $\ge$ 16 are plotted in subsequent figures, where TS is the \fermi\ test statistic and  $\sqrt{TS}$ is roughly equivalent to the detection significance per integrated bin \citep{Mattox96, Abdo09}. When daily binned fluxes fell below the significance threshold, we plot weekly binned flux, also at the TS$\ge$ 16 level. In the case of PKS~0454-234, adaptive binning techniques \citep{Lott12} were used to determine the \fermi\ \g-ray flux in the 10 day span around MJD 56280. The same TS threshold that was used to plot significant \fermi\ \g-ray fluxes during the daily and weekly analysis.

We note that simultaneous measurements of \fermi\ and Steward data means data obtained within the same day. The \smarts\ optical spectroscopy was flux-calibrated to OIR photometry to within the hour. In no case are any two datasets matched with temporal separation of more than 4 days.
\footnotetext[4]{http://fermi.gsfc.nasa.gov/ssc/data/access}

\begin{deluxetable*}{llrcr}

\tablecaption{SaMOSA Observation Summary \label{tab:samshort}}

\tablehead{\colhead{Source Name} & \colhead{\fermi\ Identifier} & \colhead{Redshift} &\colhead{No. of Observations} & \colhead{Observed Line(s)}}

\startdata
PKS 0208-512 & 2FGL J0210.7-5102  & 1.003 & 37 &  \mgii, \ciii \\
PKS 0402-362 & 2FGL J0403.9-3604 & 1.423  &9 &  \mgii, \siii, \civ \\
PKS 0454-234 & 2FGL J0457.0-2325 & 1.003 & 43 &  \mgii \\
3C 454.3 & 2FGL J2253.9+1609 & 0.859 & 35 & \mgii, H$\gamma$, H$\beta$, H$\alpha$\\
PKS 1510-089 & 2FGL J1512.8-0906 & 0.36 & 102 &  \mgii, H$\gamma$, H$\beta$, H$\alpha^\dag$ \\
PKS 2052-474 & 2FGL J2056.2-4715 & 1.489 &  8 &  \mgii, \ciii, \civ \\
PKS 2142-75 & 2FGL J2147.4-7534 & 1.139 & 17 &  \mgii
\enddata
\tablecomments{$^\dag$An emission line that may have second order contamination. See text for details.}
\end{deluxetable*}

\setlength{\tabcolsep}{3pt}
\begin{deluxetable*}{cccccccccccccccccc}
\tablecaption{SaMOSA Optical Photometry and Spectroscopy Log \label{tab:samobs}}
\tablehead{ \colhead{UTC$^\dag$} & \colhead{MJD} & \colhead{B} & \colhead{$\sigma_B$} & \colhead{V} &  \colhead{$\sigma_V$} & \colhead{R} & \colhead{$\sigma_R$} & \colhead{J} & \colhead{$\sigma_J$} & \colhead{W(\mgii)} & \colhead{$\sigma_{W(\mgii)}$} & \colhead{W(\ciii)} & \colhead{$\sigma_{W(\ciii)}$} }
\startdata
\multicolumn{18}{c}{PKS 0208-512}\\ 
\hline \\
20080623 &54640.379& 16.472&0.004&-&-&15.49&0.004&-&-&4.953&0.467&9.859&1.672& & & &  \\
20080805 &54683.331& 18.016&0.015&17.497&0.018&17.137&0.018&15.77&0.023&-&-&13.321&4.131& & & &  \\
20080823 &54701.3&18.155&0.024&17.654&0.025&17.282&0.022&-&-&18.807&2.828&16.451&3.933& & & &  \\
20080908 &54717.236&18.372&0.045&17.654&0.058&17.433&0.048&-&-&23.452&3.344&27.091&6.037& & & &  \\
20080926 &54735.151&16.738&0.007&16.283&0.008&15.882&0.007&14.494&0.011&6.5&0.927&8.273&1.567& & & &  \\
...
\enddata
\tablecomments{$^\dag$UTC is in YYYYMMDD format. The equivalent widths of the emission lines are reported as W(species), e.g. W(\mgii), in units of Angstroms. \smarts\ photometry are given in magnitudes. Table~\ref{tab:samobs} is published in its entirety in The Astrophysical Journal, a portion is shown here for guidance regarding its form and content.}
\end{deluxetable*}

\begin{figure}
\epsscale{1.2}
\plotone{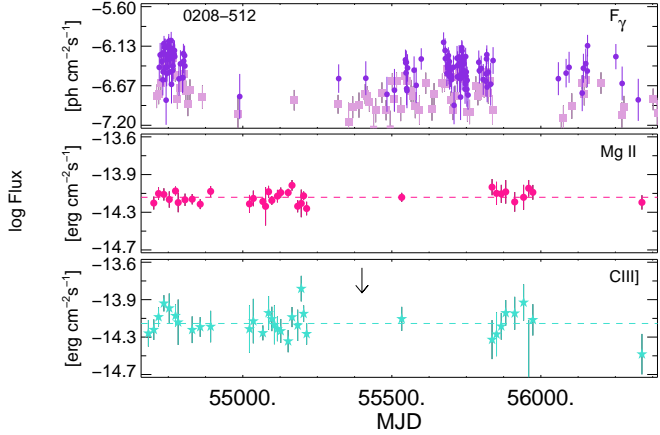}
\vspace{-17.0em}
\caption{The daily binned \fermi\ (E$>$100MeV) \g-ray photon flux (purple circles), or, when undetected in daily binning, weekly binned photon flux (light purple squares), as well as emission line fluxes of \mgii\ (magenta circles) and \ciii\ (cyan stars) for \a0208. \label{fig:elc1}}
\end{figure}

\begin{figure}
\epsscale{1.2}
\plotone{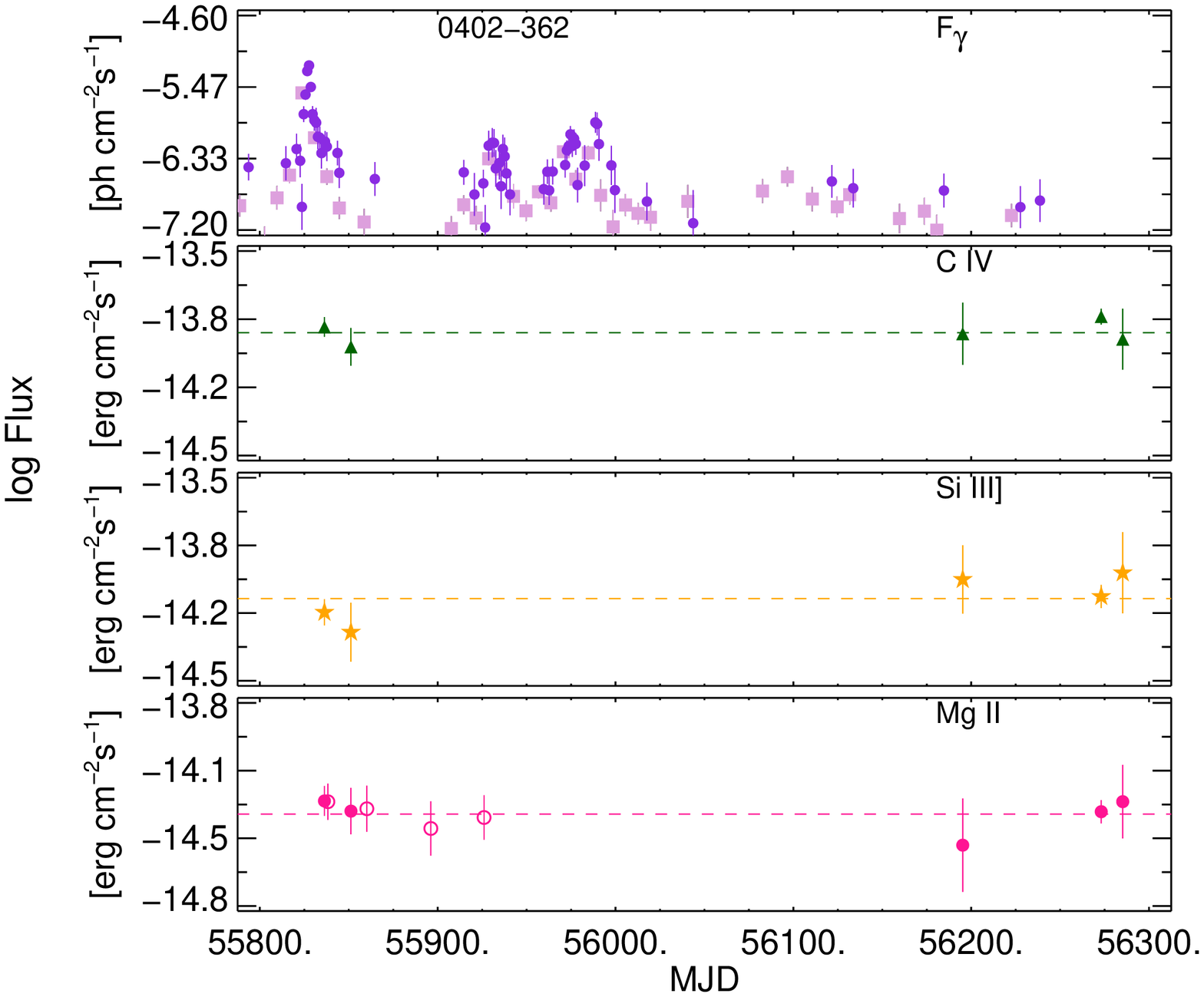}
\vspace{-11.0em}
\caption{Light curves for \ba0402. Symbols as in Figure~\ref{fig:elc1}, plus  \civ\ (green triangles), \siii\ (orange stars). Steward Observatory data are also presented in the \mgii\ light curve (open pink circles),  offset by -0.1203 dex to match the \smarts\ mean line flux measurement. As the total variability (deviation from the respective mean) is evaluated, the normalization does not impact the results. The large \fermi\ \g-ray flare at approximately MJD 55830 is not well sampled in the emission line light curve.\label{fig:elc2}}
\end{figure}

\begin{figure}
\epsscale{1.2}
\plotone{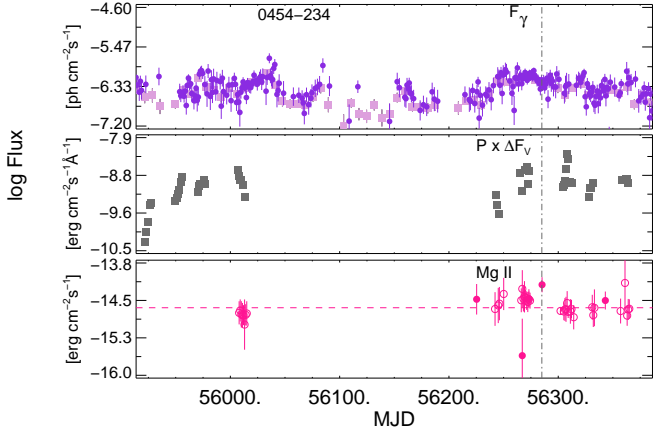}
\vspace{-17.0em}
\caption{Light curves for \ca0454, with symbols as in Figure 1. Optical linear polarimetry from Steward Observatory (grey squares), reflects the non-thermal (rather than total) flux contribution. A significant emission line flare in \mgii, peaked at log F$_{\mgii}$~=~-14.23~\fflux\ on MJD 56285, is indicated by the vertical grey dot-dashed line.  Emission line fluxes from Steward Observatory are represented by open circles and offset by --0.2493 dex. \label{fig:elc3}}
\end{figure}

\begin{figure}
\epsscale{0.8}
\plotone{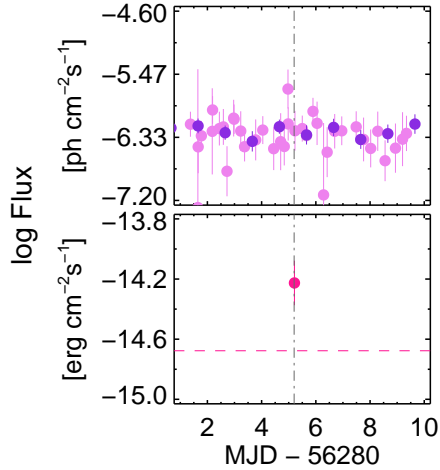}
\vspace{-35.0em}
\caption{\fermi\ \g-ray light curves for \ca0454, centered on the significant \mgii\ emission line flare. Hourly bins (lavender circles; top), derived using the adaptive binning technique \citep{Lott12}.  The fluxes produced via the adaptive binning method meet the same TS threshold as the daily binned fluxes (TS $\ge$ 16).  \label{fig:elc3f}}
\end{figure}

\begin{figure*}
\plotone{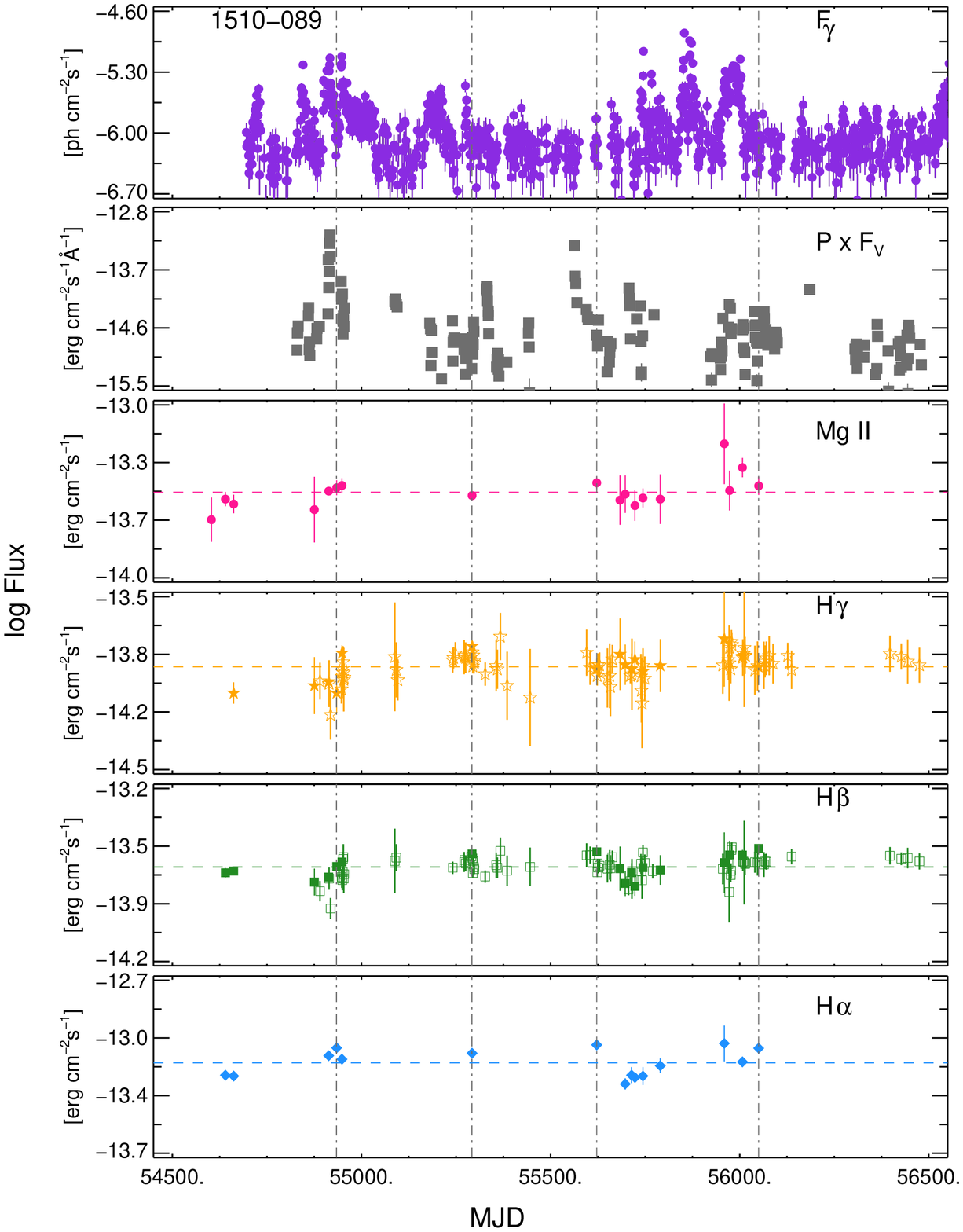}
\caption{Light curves for \df1510. Symbols as in previous Figures,  plus \hg\ (orange stars), \hb\ (green squares) and \ha\ (blue diamonds). Four well-defined emission line flares are observed, as indicated by the grey dot-dashed lines. In the \fermi\ \g-ray flaring period from MJD~554850~-~55000, \ha\ underwent a strong line flare that peaked on MJD~54934 at F$_{H\alpha}$~=~-13.97~\fflux. During the same \g-ray flare, TeV photons were detected by H.E.S.S. between MJD~54910~-~55952. \hg\ and \hb\ show line flares on MJD~55292 at 3.3$\sigma$ and 3.7$\sigma$, respectively. \label{fig:elc4}}
\end{figure*}

\begin{figure*}
\plotone{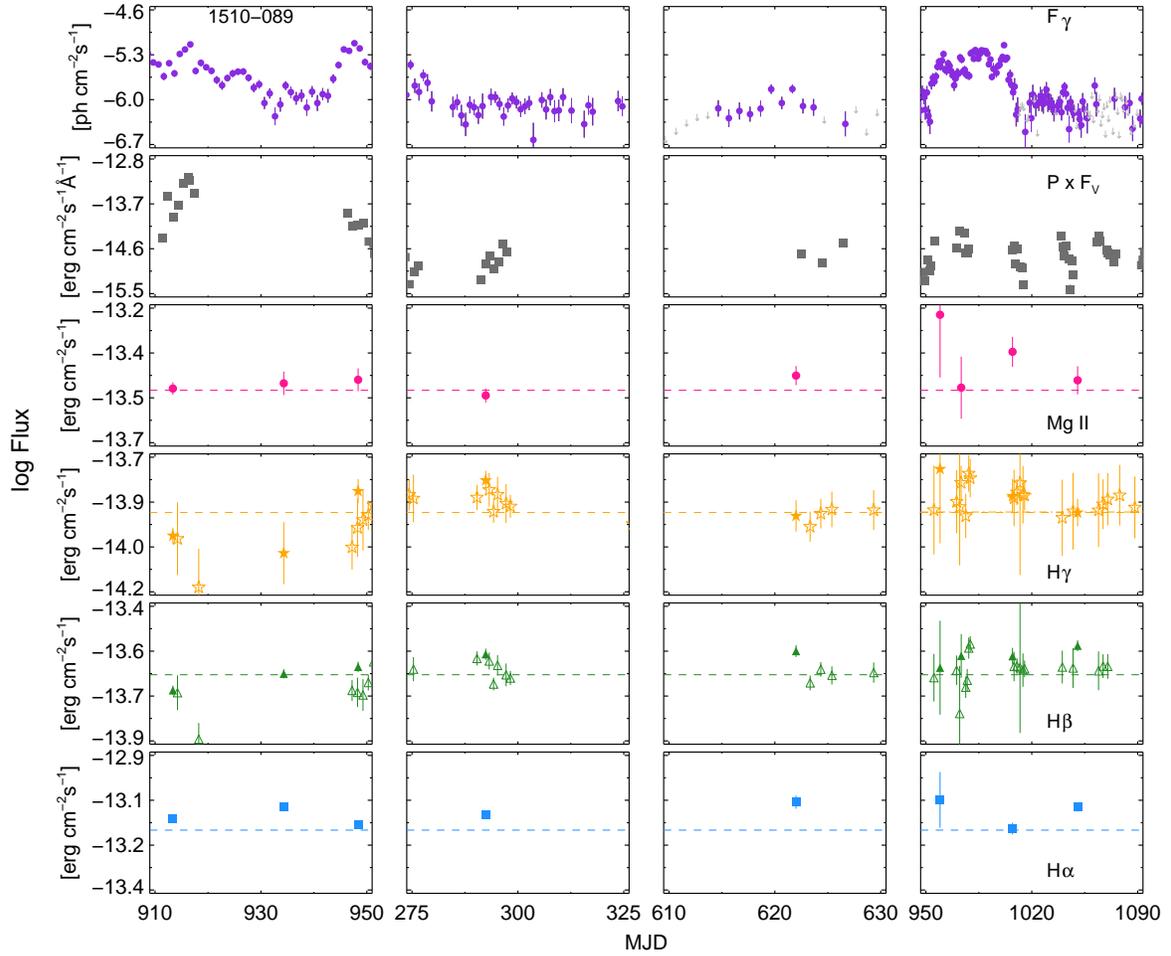}
\caption{The light curve for \df1510, marked as in Figure~\ref{fig:elc4}, now shown with the four significant emission line flaring periods isolated for each emission line. Dates are given in units of MJD - 55000 in all but the first panel, which are given in MJD - 54000. In each case, there is a \fermi\ \g-ray flare (or increased \g-ray emission) associated with each emission line flare. However, not every emission line had detectable emission line variability.  \label{fig:4f}}
\end{figure*}

\begin{figure}
\epsscale{1.2}
\plotone{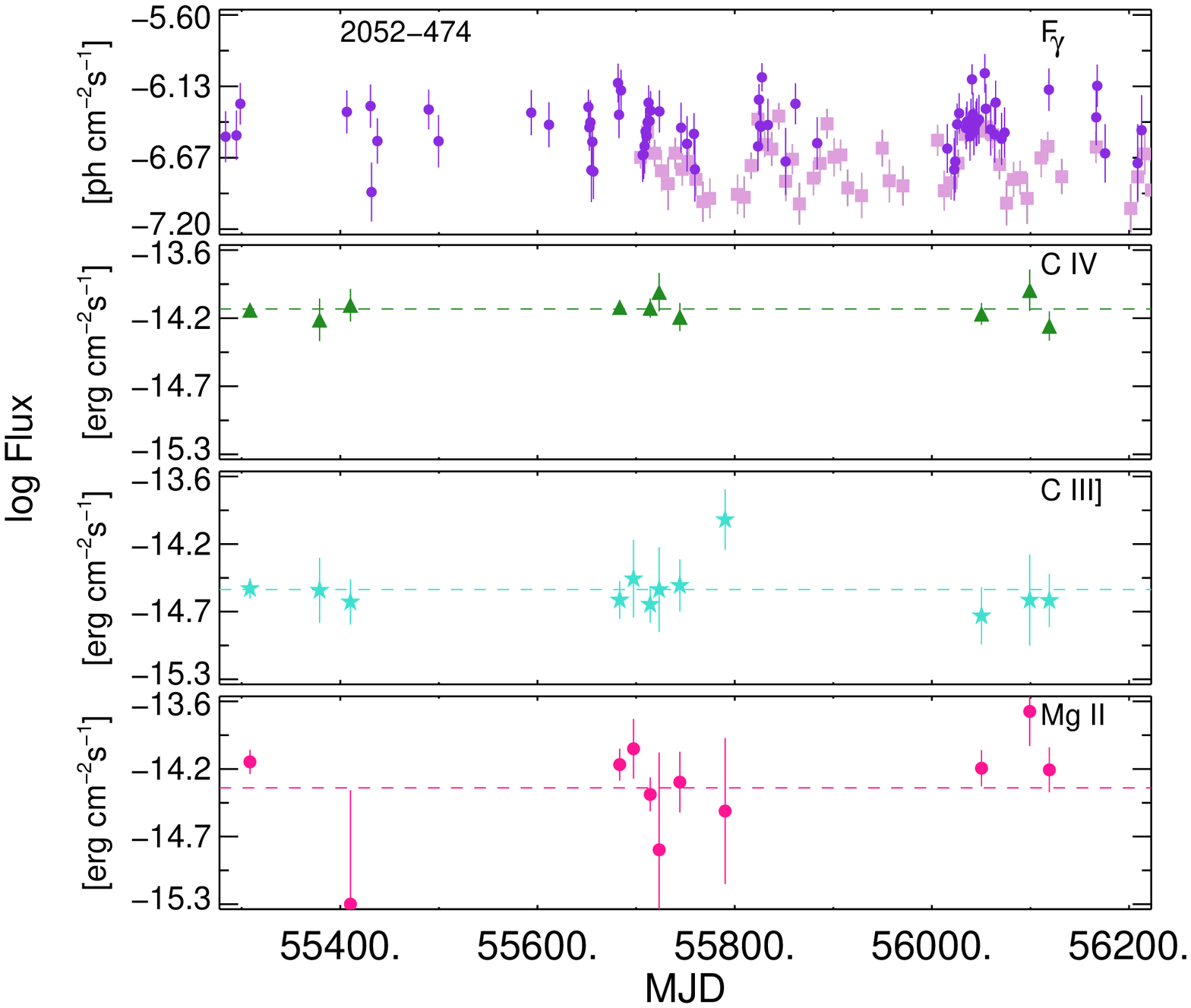}
\vspace{-11.0em}
\caption{Light curves for \e2052. Symbols as in previous Figures.  \label{fig:elc5}}
\end{figure}

\begin{figure}
\epsscale{1.2}
\plotone{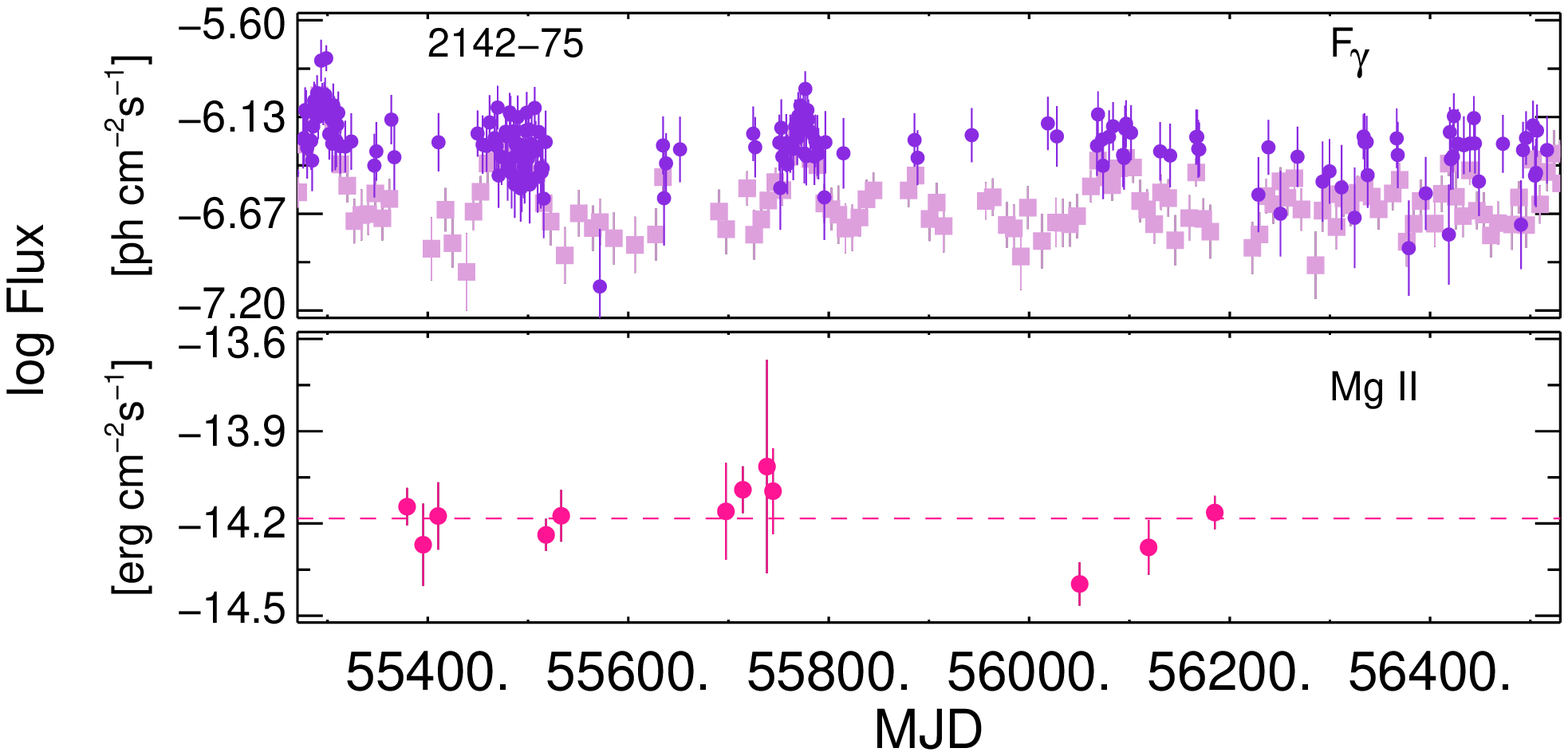}
\vspace{-22.0em}
\caption{Light curves for \f2142. Symbols as in previous Figures. \label{fig:elc6}}
\end{figure}

\begin{figure}
\plotone{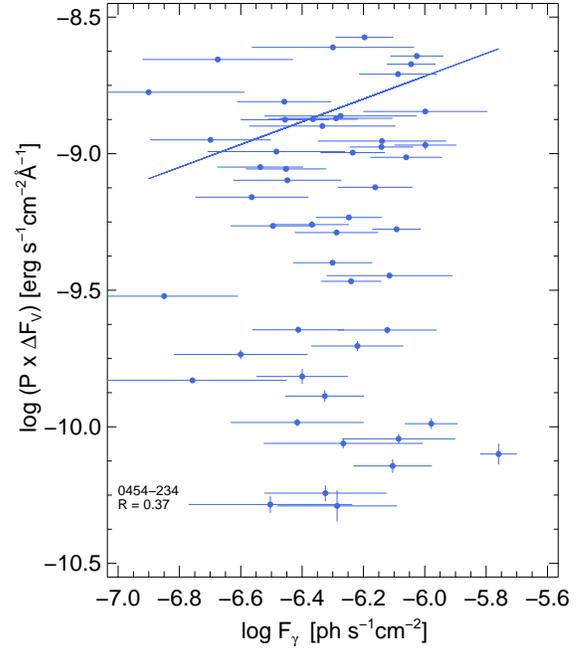}
\vspace{1.0em}
\caption{Differential optical polarized flux vs. \fermi\ \g-ray flux for \ca0454. The correlation coefficient for the linear regression is 0.37. Polarization is not a ``clean'' measure of the strength of the non-thermal emission, given its dependence on both the order of the magnetic field strength and the source brightness. For this source, the non-thermal synchrotron flux present in the optical spectrum is essentially unrelated to the intensity of the \g-ray flux. \label{fig:polgam2}}
\end{figure}

\begin{figure}
\plotone{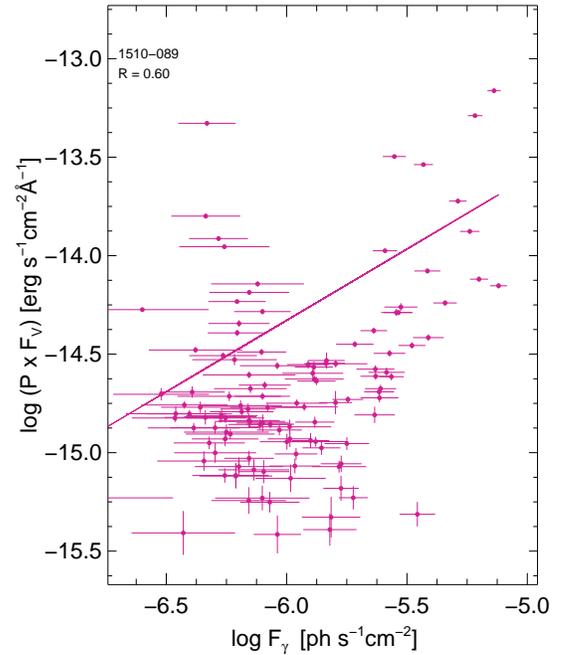}
\vspace{1.0em}
\caption{Optical polarized flux vs. \fermi\ \g-ray flux for \df1510. The correlation coefficient for the linear regression is 0.60. The non-thermal synchrotron flux in the optical spectrum is not very strongly correlated with the \g-ray flux but a trend is clear. \label{fig:polgam1}}
\end{figure}

\begin{figure}
\plotone{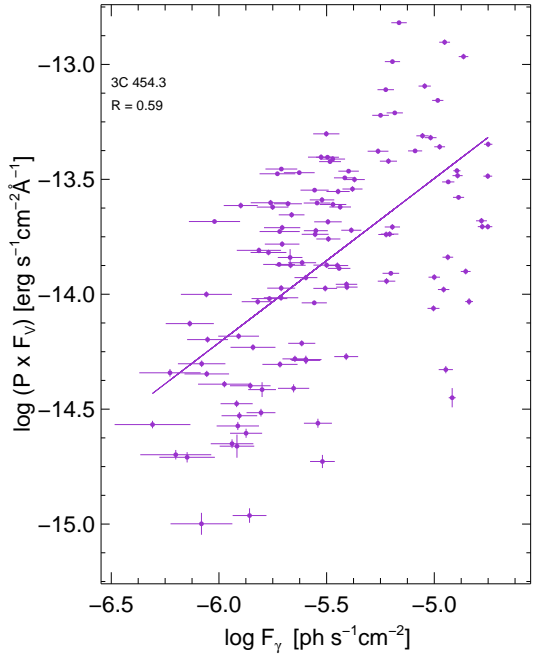}
\vspace{1.0em}
\caption{Optical polarized flux vs. \fermi\ \g-ray flux for \thesource. The correlation coefficient for the linear regression is 0.59. Although the correlation is weak, we see that the polarization does increase with increasing \g-ray flux. \label{fig:polgam3}}
\end{figure}

\begin{deluxetable}{llccc}
\tablecaption{Mean Emission Line and \g-ray Fluxes\\ and Luminosities\label{tab:sammean}}
\tablehead{\colhead{Source} & \colhead{Line} & \colhead{$\langle$F$_{line}\rangle$}& \colhead{$\langle$L$_{line}\rangle$}& \colhead{$\langle$L$_\gamma\rangle$}}
\startdata
PKS 0208-512 & Mg II & -14.17 (0.06) & 43.94& 51.29 (0.20) \\
             & C III] & -14.18 (0.18) & 43.95 & \nodata \\
PKS 0402-362 & C IV  & -13.90 (0.06) & 44.20 & 51.74 (0.28) \\
	     & S III] & -14.10 (0.12) & 44.01 & \nodata   \\
	     & Mg II$^\dag$ & -14.34 (0.08) & 43.38 & \nodata \\
PKS 0454-234 & Mg II$^\dag$ & -14.64 (0.07) & 43.08  & 51.35 (0.24) \\
PKS 1510-089 & Mg II & -13.51 (0.10) & 43.14 & 50.62 (0.31)\\
	     & H$\gamma^\dag$ & -13.92 (0.19) & 42.72 & \nodata \\
	     & H$\beta^\dag$  & -13.65 (0.20) & 42.97 &\nodata \\
             & H$\alpha$ & -13.20 (0.08) & 43.44 & \nodata \\
PKS 2052-474 & Mg II & -14.09 (0.09) & 43.48 & 51.75 (0.19) \\
	     & C IV  & -14.55 (0.20) & 44.07 & \nodata \\
	     & C III] & -14.82 (0.44) & 43.67 & \nodata \\
PKS 2142-75  & Mg II & -14.18 (0.10) & 43.69 & 51.58 (0.20)\\
3C 454.3 & Mg II & -13.96 (0.10) & 43.60 & 52.17 (0.39)\\
& H$\beta$& -13.93 (0.14) & 43.61 & \nodata \\
& H$\gamma$& -13.87 (0.14) & 43.67 & \nodata
\enddata
\tablecomments{Mean line flux and line luminosity is given in units of log \fflux\ and log erg s$^{-1}$, respectively, and includes all available data. $^\dag$The systematic offset between the Steward and \smarts\ data has been applied before calculating the mean line fluxes and luminosities. \fermi\ \g-ray mean luminosity includes all \fermi\ data with TS $\ge$ 16 in the entire observing window for each source and is in units of log erg s$^{-1}$. The standard deviation is the reported uncertainty. Data for \thesource\ were previously reported in \citet{Isler13}, but are reproduced here for comparison.}
\end{deluxetable}

\begin{center}
\begin{deluxetable}{llccccc}
\setlength{\tabcolsep}{0.02in}
\footnotesize
\tablecaption{Significant Emission Line Flares\label{tab:samflares}}
\tablehead{\colhead{Source}       & \colhead{Line}      & \colhead{MJD$^\dag$}  & \colhead{UTC$^\ddagger$} & \colhead{Line Flux$^\dag$} & \colhead{$\sigma_{line}$} & \colhead{S$^\ast$ } } 
\startdata
\ca0454\ & \mgii\      &  56285 & 20121222 & -14.23 & 0.15 & 3.0 \\
\df1510\	     & H$\alpha$ &  54934  & 20090411 & -13.09 & 0.01 & 8.5 \\
                     & H$\alpha$ &  55292 & 20100404 & -13.12 & 0.02 & 3.8 \\
	     & H$\gamma$ & 55292  & 20100404 & -13.79& 0.04 & 3.3 \\
	     & H$\beta$  & 55292  & 20100404 & -13.58 & 0.02 & 3.7 \\
	     & H$\beta$  & 55622  & 20110430 & -13.57 & 0.02 & 4.0 \\
	     & H$\beta$ & 56050 &20120501 & -13.55 & 0.02 & 5.4 \\
\thesource**   & \mgii\ & 55165 & 20091130 & -13.73 & 0.06 & 1.8\\
                     & H$\gamma$ & 55165 & 20091130 & -13.58 & 0.09 & 2.8\\
                     & H$\gamma$ & 55518 & 20101118 & -13.43 & 0.06 & 3.7
\enddata
\tablecomments{Emission line fluxes are in units of log \fflux. $^\dag$The MJD and associated emission line flux is given for the peak value. $^\ddagger$UTC is given in YYYYMMDD format. $^\ast$Detection significance, S, is given in units of $\sigma$ away from the mean line flux. **The significances for \thesource\ are reproduced here for comparison.}
\end{deluxetable}
\end{center}

\begin{deluxetable}{llcccc}
\tablecaption{$\chi^2$ Variability Statistics \label{tab:chisq}}
\setlength{\tabcolsep}{0.01in}
\tablehead{\colhead{Source} & \colhead{Line} & \colhead{DOF$^\dag$} & \colhead{$\chi^{2}$} & \colhead{Prob.}  & \colhead{$\sigma$}}
\startdata
PKS 0208-512 & Mg II              & 33 & 27.8  & 0.72 & 0.35 \\ 
                       & C III]                & 34 & 59.6  & 0.004 &  2.86 \\
PKS 0402-362 & Mg II              & 4    & 1.39 & 0.84 & 0.20 \\ 
                      & Si III               & 4    & 3.13 & 0.54 & 0.62\\
                       & C IV                & 4    & 5.03& 0.28 & 1.07\\
PKS 0454-234 & Mg II               & 4   & 12.9 & 0.01 & 2.51 \\ 
PKS 1510-089 & Mg II               & 17 & 18.4 & 0.36  & 0.91 \\ 
                        & H $\gamma$ & 15 & 32.3 & 0.01  & 2.76 \\
                        & H $\beta$      & 17 & 87.9 & $\ll 10^{-3}$ & 6.75\\
                        & H $\alpha$    & 14 & 199.2 &$\ll 10^{-3}$  & 12.3\\
PKS 2052-474    & Mg II              & 9     & 17.0         &0.07 & 1.78\\
	        & C III]                & 11    &7.78          & 0.73        & 0.34\\
	        & C IV                & 9     & 3.94         & 0.91      & 0.11\\
PKS 2142-75 & Mg II                 & 11     & 14.5 &  0.21 & 1.26\\
3C 454.3     & Mg II & 24 & 61.9 & $\ll 10^{-3}$  & 4.14\\
 	& H$\beta$ & 25 & 86.9 & $\ll 10^{-3}$ & 5.64 \\
	& H$\gamma$ & 26 & 43.2 & 0.02 & 2.36
\enddata
\tablecomments{$^\dag$ DOF are the degrees of freedom. $\chi^{2}$ is the total statistic. Prob. is the probability that the flux deviates from the null result and $\sigma$ is the significance from the mean, assuming a normal distribution.}
\end{deluxetable}

\section{Results}
\label{sec:samres}

We evaluate the emission line variability of the SaMOSA sample by two independent methods. First, we define an empirical emission line flare as a significant excursion of the line flux above the mean level, following a prescription similar to that presented in \citet{Nalewajko13}, but adapted to the present dataset. Specifically, we define an empirical emission line flare if 1) at least three consecutive points, with any two sequential points separated by $\leq$ 60 days, and 2) at least one point between the \textit{first} point to deviate from the mean and the \textit{last} point to deviate from the mean must be $\ge$ 3$\sigma$ above the mean line flux. The empirical line flares are recorded in Table~\ref{tab:samflares}. 

The second method for evaluating emission line variability is the $\chi^2$ variability test. We compute the $\chi^2$ statistic for an assumed constant line flux (the null hypothesis) and use the number of degrees of freedom (DOF) to calculate the probability that the given $\chi^2$ deviates from the null hypothesis. Unless otherwise noted for an individual source, the $\chi^2$ variability test was consistent with the null hypothesis at the p$>$ 0.05 level. Results can be seen in Table~\ref{tab:chisq}. This method is sensitive to the calculated uncertainty in the emission lines. The \smarts\ spectra have better-constrained uncertainties than the Steward data, which were derived \textit{a posteriori}, so while any variability detected in the \smarts\ data using this method was not negated by the addition of Steward data, the statistical significance was weakened in some cases due to the larger uncertainty on a given line flux measurement. For this reason, we calculate the $\chi^2$ statistics based only on the \smarts\ data, for which the uncertainties are well-characterized.

Figures~\ref{fig:elc1}-\ref{fig:elc6} show the \fermi\ \g-ray and emission line light curves for the SaMOSA sample.

\begin{center}
\textit{Notes on Individual Sources}
\end{center}

\subsection{\a0208}
\a0208\ was observed with \smarts\ from 2008 August 6 - 2013 February 12 (MJD~54640~-~56339) and the emission line behavior is shown in Figure~\ref{fig:elc1}.  Although a slight increase in line flux was detected in \ciii\ between MJD~54701~-~54892, peaking at MJD~54735, with log F$_{\ciii}$~=~-13.97 \fflux\ (2.5$\sigma$), it does not meet the empirical emission line flare significance criteria for an empirical emission line flare and is not included in subsequent analysis. Thus, no significant emission line variability was seen in this blazar.

\subsection{\ba0402}
\ba0402\ was observed from 2011 October 2 - 2012 December 12 (MJD~55836~-~56285). The source underwent a large, short-duration \g-ray flare from MJD~55821~-~55838 that is not well-sampled in emission line flux (see Figure~\ref{fig:elc2}). No significant emission line variability was detected in this blazar.

\subsection{\ca0454}
\ca0454 was observed between 2011 October 16 - 2013 July 27 (MJD~55850~-~56500); the light curve can be seen in Figure~\ref{fig:elc3}. We also plot the optical linear polarization. For this object, calibrated V-band photometry from Steward Observatory was not available, so we used differential photometry ($\Delta$F$_V$) to determine the relative polarized flux. 
 \mgii\ underwent a 3.0$\sigma$ line flare with peak line flux log~F$_{Mg II}$~=~-14.23 \fflux\ on MJD 56285. No accompanying \fermi\ \g-ray flare was detected in the daily binned fluxes, so the adaptive binning technique \citep{Lott12} was utilized during the 10-day period around MJD 56280 to determine if sub-day variability was present. We plot the results of the adaptively binned \fermi\ \g-ray fluxes along with the daily binned \g-ray fluxes and the \mgii\ emission line flare for comparison in  Figure~\ref{fig:elc3f}. The apparent deviation of \g-ray flux derived from adaptive binning is of the same order as the textit{rms} in this region and is thus not significant.

\subsection{\thesource}
We first presented the emission line light curve for \thesource\ in \citet{Isler13}; one can also be seen in \citet{LT13} for this epoch. We extend the previous analyses by applying the criterion for emission line flares used here; both previously reported emission line flares in \hg\ and \mgii\ meet both variability criterion and are thus statistically significant. 

\subsection{\df1510}
Optical spectra were obtained for \df1510\  between 2008 May 17 - 2012 May 3 (MJD~54603~-~56050); the light curve can be seen in Figure~\ref{fig:elc4}. We detected line flares in \ha\ peaking on MJD 54934 at log~F$_{\ha}$=~-13.09 \fflux\ with 8.5$\sigma$ significance.  The leading line flux for this \ha\ flare was also above the mean at MJD 54913 with 3.3$\sigma$ significance (log~F$_{\ha}$=~-13.14 \fflux), suggesting that the emission line flare may extend past the range observed on the increasing side of the \g-ray flare. During the \fermi\ \g-ray flare (MJD~54850~-~55000) with peak flux F$_\gamma$~=~--5.08 \pflux\ on MJD 54917,  H.E.S.S. also detected very high energy photons from MJD~55910~-~55952, with highest emission, log F$_{>0.15 TeV}$~$\approx$~-10.4 \pflux\ on MJD 54915 \citep{Hess13}. 

\hg\ and \hb\ show significant line flares on MJD 55292 at the 3.3$\sigma$ (log~F$_{\hg}$~=~-13.79 \fflux) and 3.7$\sigma$ (log~F$_{\hb}$=~-13.58 \fflux) level, respectively. A significant emission line flare was observed in \hb\ on MJD~55622 with  log~F$_{\hb}$=~-13.57 \fflux\ (4.0$\sigma$). \hb\ has an emission line flare during the same \g-ray flare, peaking on MJD 56050 at log~F$_{\ha}$=~-13.54 \fflux (5.4$\sigma$). In Figure~\ref{fig:4f} we show the regions where emission line flares were detected in at least one emission line, as described above.

Statistically significant emission line variability in the \ha, \hg\ and \hb\ emission lines is measured and the $\chi^2$ variability test also indicates variability: $\chi^2_{\ha}$~=~199.4 (14 degrees of freedom, p$\ll$ 10$^{-3}$), $\chi^2_{\hg}$~=~32.3 (15 degrees of freedom, p=0.01) and $\chi^2_{\hb}$=~87.9 (17 degrees of freedom, p$\ll$10$^{-3}$).

\subsection{\e2052}
 \e2052\  was observed from 2011 May 2 - 2012 July 11 (MJD 55683 - 56119) and is the only source in the sample that did not show any significant \fermi\ \g-ray flux above log F$_\gamma$~=~-6~\pflux\ over the epoch of observation presented here (see Figure~\ref{fig:elc5}). No significant emission line variability was observed in this blazar.

\subsection{\f2142}
\f2142\ was observed from 2010 May 5 - 2012 September 15 (MJD 55321 - 56185), seen in Figure~ \ref{fig:elc6}.  No significant emission line flares were observed during the epoch of observation.

\begin{deluxetable}{cc}
\tablecaption{\fermi\  2FGL Variability Index}
\tablehead{\colhead{Source Name} & \colhead{Variabiity Index}}
\startdata
\thesource\ & 14189\\
\df1510\ & 6405\\
\ca0454\ & 1501\\
\ba0402\ & 1417\\
\f2142\ & 1162\\
\e2052\ & 791\\
\a0208\ & 733
\enddata
\tablecomments{The variability index is obtained from the \fermi\ 2 yr Source Catalog \citep{Nolan12}. \label{tab:vi}}
\end{deluxetable}

\section{Accretion Disk - Jet Interaction}\label{sec:samcols}
The statistically significant emission line variability seen in \ca0454, \thesource\ and \df1510, but not in less \g-ray active sources, suggests that in the most active \g-ray sources the broad line region may be partially photoionized by the jet. In this case, a correlation, potentially with lags, between the \g-ray flux and emission line flux is expected. We characterize the \g-ray jet activity by using the \textit{variability index}, as defined in the second \fermi\ catalog \citep[2FGL;][]{Nolan12}, as the sum of 2 $\times$ log(likelihood) comparison between the flux fitted in 24 time segments and a flat light curve over the full 2 yr catalog interval. Values greater than 41.64 indicate that there is less than 1\% chance of being a steady source. We find that the three blazars in which we identified emission line variability in this sample also have the highest variability index, although all the sources in this sample are consistent with \g-ray variability with high certainty. We list the variability indices in Table~\ref{tab:vi}.

A discrete correlation analysis would be useful to test such a correlation, however, with the small number of data points in a given emission line light curve it is not valid for this data set. We attempted to quantify the significance of the correlation between \fermi\ \g-ray flare and emission line flare above a random occurrence by carrying out a Monte Carlo simulation. We ran 1000 iterations of randomized emission line fluxes with respect to the date of spectroscopic observation and then applied our two tests of emission line variability to each iteration. Because the distribution of the randomized emission line fluxes is completely dominated by clustering due to the seasonal observation schedule and relatively small number of observations,  we were not able to derive a meaningful significance.

If this explanation of a possible correlation (potentially with temporal offset) between the \g-ray flux and emission line flux is correct, then other empirical indicators should confirm non-thermal emission in the optical regime during emission line flares. To test the presence of such emission, we consider the optical polarization of \ca0454, \thesource\ and \df1510. Polarization data are not available for the other sources in the sample, as they are too far south of Steward Observatory.

While optical photometry measures the total emission from both the accretion disk and the jet, the optical polarization measures the contribution of synchrotron flux in the optical-ultraviolet regime \citep{Raiteri12, Smith94}. Thus, by calculating the polarized flux, we can measure the level of non-thermal emission at the same time as the \g-ray and emission line fluxes in order to distinguish between the thermal contribution and additional emission from the jet. However, the polarized flux depends on both the magnetic field ordering of the synchrotron region, and how bright it is. These two parameters are not often closely related, and as a result, the polarized flux is not a ``clean'' measure of the strength of the non-thermal continuum compared to the \g-ray flux (which comes only from the jet). Yet, we can associate this emission with non-thermal emission given the featureless spectrum and rapid variability of the polarized flux, which is too rapid to originate from scattered accretion disk thermal emission.

We plot the polarized flux light curve for \ca0454\ and \df1510\ in Figures~\ref{fig:elc3} and \ref{fig:elc4}, respectively, and refer the reader to \citet{Isler13} for a similar plot for \thesource. Visual inspection confirms the general agreement of an increase in polarized flux during \g-ray flares, although temporal offsets can be seen between bands. We plot the polarization with respect to \g-ray flux of these 3 blazars in Figures~\ref{fig:polgam2}, \ref{fig:polgam1} and \ref{fig:polgam3}. We find a correlation coefficient of R = 0.37, R = 0.59 and R = 0.60 for \ca0454, \thesource\ and \df1510, respectively.  Thus, we suggest that there is an increase in polarized light during periods of increased \g-ray activity in \thesource\ and \df1510, while we do not infer such a relationship in \ca0454\ based on these data. 

Taken together, the polarization, \g-ray and line flux diagrams provide tentative evidence that, during \g-ray flaring events, an additional population of non-thermal ionizing photons produce enough photoionizing flux to increase the emission line flux. We infer this from the temporal proximity (but not necessarily simultaneity) of the emission line fluxes to \g-ray flares in the light curve and the coincidence of optical polarization during periods of high \g-ray flaring and emission line fluxes. We note that all of the proxies for correlation here are likely diluted by offsets in the timescales of increasing flux in the different bands. Still, these empirical results suggest that there could be a significant source of photoionizing flux being produced by a non-thermal jet, and that this emission may be contributing to the increase in emission line flux that we observe. 

\section{Discussion}\label{sec:samdisc}

\subsection{Emission Line Variability}
We find that 4 of the 7 blazars observed in this sample show no evidence of statistically significant emission line variability. The lack of emission line variability in these blazars is consistent with the standard model that the accretion disk is the predominant source of photoionizing flux. 

However, \ca0454, \thesource\ and \df1510\ all show statistically significant emission line variability, with a mean peak emission line flare significance of 4$\sigma$.We are only aware of a few blazars with published simultaneous \fermi\ \g-ray and optical emission line data. Simultaneous emission line variability studies in \fermi-monitored blazars have generated mixed results. Among a set of similar \g-ray and optically bright, variable quasars, little emission line variability was detected in PKS~1222+216 or 4C 38.51 \citep{Smith11, Farina12, Raiteri12}. It was reported that FSRQ PKS~1222+216 did not have significant emission line variability during recent \fermi\ \g-ray flaring events \citep{Smith11, Farina12}, however line variability at the 2.6$\sigma$ level is evident in both datasets. While emission line variability at this level would not have met the criteria set forth in this work, the lines do show some variation based solely on continuum variations. Without reanalyzing the data, we note that when the H\b\ line luminosity from Steward Observatory reported in \citet{Farina12} is compared to the line luminosity obtained from TNG in the same work, there is a 17$\sigma$ difference in line luminosity. The latter was obtained following a \textit{MAGIC} triggered observation.

\subsection{Correlated Variability}
Emission line variability has been detected in \ca0454, \thesource\ and \df1510, which are also the three sources with the highest \fermi\ variability index. This is suggested by the empirical emission line flares and $\chi^2$ variability test, correlation information including possible lags is harder to assess given the limits of the data. Thus, we attempt to relate the degree of non-thermal jet contribution to the photoionizing flux in the broad line region to optical polarization. It has long been known that during periods of increased \g-ray activity, the thermal contribution to the optical-ultraviolet continuum is swamped by non-thermal emission \citep[e.g.,][]{Smith94}. We confirm this result with the optical polarization for \thesource\ and \df1510, where brighter \g-ray (and emission line) fluxes correspond to highly polarized states, presumably due to jet emission. However, we do not find evidence of this behavior in \ca0454. The synchrotron peak has been shown to be well-correlated from infrared to ultraviolet for FSRQs \citep[e.g.][]{Bonning09}, such that optical variations indicate similar variability patterns in the ultraviolet, where the photoionizing flux peaks.  

The tentative picture that emerges from this work is that the most active \g-ray flaring sources have evidence of emission line variability that is not seen in sources with less active jets. The correlations between \g-ray flux and emission line flux are likely diluted by the presence of lags and/or leads in the peak of either curve. Thus, we expect that correlated line variability could be detected with higher cadence, multi-epoch optical emission line studies of \g-ray active blazars.

\subsection{Location of the \g-emitting Region}
Next, we turn our attention to the location of the \g-emitting region, which has been the subject of much research. The models fall into two major categories: 1) near-field \g-emitting models suggest that the bulk of the jet dissipation occurs on sub-pc scales, either very deep in the Broad Line Region \citep[BLR;][]{Poutanen10} or near the edge of the broad line region \citep{Bottcher07, Kataoka08, Tavecchio10, Ghisellini10, Poutanen10, Tavecchio10, Stern11, Abdo11}, but in any case at or below canonical distances of 0.1~pc \citep[e.g.][]{Peterson93, Peterson06}, and 2) far-field \g-emitting scenarios, which suggest jet dissipation on much larger spacial scales (tens of pc) from the central source \citep{Marscher11, Agudo11, Jorstad12}. These studies attempt to constrain the location of the \g-emitting region based on SED modeling, correlation studies, and/or ultra-short \g-ray variability. 

More recently, a few coordinated optical spectroscopic variability studies, like the one presented here, have been undertaken in an effort to identify emission line variability of \g-ray bright blazars \citep{Smith09, Benitez10, Raiteri12, Smith11, Isler13, LT13}. In the cases where line variability is found, attempts are sought to directly (and simultaneously) relate the broad line variability to jet variability via \g-ray, mm or other non-thermal emission. 

The results of both the indirect and direct studies have been mixed, even for the same source and same flare. For example, in \thesource, following the 2009 December and 2010 November flaring periods, \citet{Tavecchio10} and \citet{Abdo11} suggest the \g-emitting region of \thesource\ could be located at the outer edges of the broad line region (r$_{em} \sim$ 0.14 pc), using \g\g-opacity arguments. \citet{Isler13} also suggested a near-field dissipation mechanism after observing statistically significant emission line variability in both the 2009 and 2010 flares. By contrast, \citet{LT13} argue in favor of a potentially far-field dissipation mechanism in the 2010 flare, given statistically significant emission line variability in close temporal proximity to a mm-core ejection. They argue that their lack of detectable emission line variability during the 2009 flare, in combination with the absence of an additional mm-core ejection, suggests that emission line variability may be caused by the radio core ejections. The two spectroscopic studies come to different conclusions likely due to different observation windows around the 2009 flare. The observations presented by \citet{LT13} did not extend across the entire \g-ray flare, therefore, a simultaneous comparison of the peak \g-ray to emission line fluxes was not possible. The data collected by \citet{Isler13} extended across the entire \g-ray flare,albeit with fewer total observations. Thus, the lack of detected emission line variability in the 2009 flare by \citet{LT13} is likely due to lack of temporal coverage and not to the absence of line variability itself.

We also consider the mm-core ejections seen in \df1510\ \citep{Marscher10} with respect to the emission line variability reported here. We report emission line variability in \ha\ which peaks on MJD~54934. A mm-core ejection is not seen in this flaring period until MJD~54959, suggesting that it is likely not the cause of the emission line variability (on the pc scales on which the core ejection is observed). To our knowledge, no subsequent studies on core ejections in \df1510\ have been published. Therefore, we are unable to compare our reports of emission line variability. While these two examples of non-coincident core ejections at the time of emission line variability do not preclude such occurrences, we suggest that a core ejection is not a necessary condition for emission line variability. However, the near temporal proximity of \g-ray flares to nearly every instance of emission line variability does suggest that jet flares (and the attendant increase in photoionizing flux) are likely required to produce emission line variability on timescales of a few weeks to months. Thus, we consider our results to favor a near-field jet dissipation region. 

\subsection{Broad Line Region Structure}
While the emission line flux variability is evidence of a BLR response, it does not provide a conclusive test of its dynamical structure. Whether the jet emitting region is within the canonical BLR or part of an outflowing wind (either from the disk or entrained in the jet) on larger scales, is still unknown. The broad emission line profiles alone are not sufficient to distinguish between broad line region dynamical models \citep{Capriotti80}, although higher line profile moments like asymmetry can distinguish steady-state versus dynamical theories \citep{Capriotti81}. Thus, whether the BLR is a distribution of gas clouds (or filaments) driven primarily by Keplerian velocities, or a disk (or jet) wind cannot be constrained by the current line variability studies; higher resolution (and cadence) coordinated spectroscopic observations are necessary to truly constrain the dynamical nature of the BLR in blazars. 

However, if some part of the BLR were located very far from the central source as a result of entrainment in the jet, we would expect to see a non-variable line core with highly variable (blueward) line wings due to the high velocities of these outflowing and entrained clouds. Conversely, we do not expect the jet emitting region to be located deep within the broad line region, as it would require a more isotropic (likely unbeamed) jet contribution that would not be significantly brighter than the accretion disk in the source frame and hence not produce detectable emission line variability. In addition, observations of very high energy (TeV) emission coincident with emission line variability, like that seen in \df1510, are hard to reconcile with the increased probability for \g\g-absorption deep in the rich BLR photon field.  

If, instead, the jet emission were interacting with the broad line region, as suggested in the `mirror model' \citep{Ghisellini96}, one could still observe jet-augmented photoionization in the broad line emission. According to this model, which estimates the BLR as a thin shell, the energy density increases dramatically at the location of the BLR, where the gas sees strongly beamed flux from the jet (see their Figure 2). Two factors are at play here: 1) the jet emission is beamed and illuminates only a small fraction of the assumed spherical BLR, within a solid angle $\pi$/$\Gamma^2$, though with a strongly enhanced flux, and 2) the disk radiation is assumed to be roughly isotropic. Together with equation 26 of \citet{Ghisellini96}, this implies the measured BLR luminosity, L$_{BLR}$ $\sim$~L$_{disk}$+$\frac{L_{jet}}{\Gamma^2}$, where both L$_{disk}$ and L$_{jet}$ are observed luminosities. For an $\Gamma \sim$~10, the jet contribution to the total (optical-UV) photoionizing luminosity is a factor of $\sim$~100 less that that of the disk. Thus, for the observed variation of a factor $\sim$~2 in BLR flux, the jet photoionizing flux should increase by a factor of $>$ 100.

For $\thesource$ and \df1510, this constraint is easily accommodated, as both have shown significant increases ($\Delta$B $\gtrsim$ 3 mag) in optical-UV flux during flaring events and L$_{jet}$ $>$ L$_{disk}$, especially in high flaring states. However, as was shown in \citep{Marscher10}, the emitting region in a given blazar is quite complicated and has been observed to shift from sub-pc to several pc scales in a matter of months \citep{Finke10, Ghisellini13} and thus we do not expect every flare to take place within the BLR. 

In the case where the \g-emitting region is located outside the BLR, we would not expect flares in the jet to cause line variability. However, lines can vary because of variable disk emission; this should happen on longer time scales (since the disk varies slowly compared to the jet) and need not, in general, be associated with an increase in \g-ray emission. The latter kind of event may have taken place in \ca0454, where we see insufficient optical-UV jet photoionizing flux combined with the lack of a strong \g-ray flare near the occurrence of the emission line flare suggest that this instance of line variability may be caused by a different physical mechanism. In this case, the accretion disk variability could be caused by hotspots which can change the shape and behavior of the optical spectrum as described by \citep{Ruan14}. Furthermore, \citet{Ghisellini10} has already shown that L$_{disk}$ $\approx$ L$_{jet}$ in \ca0454, such that the jet does not produce enough photoionizing radiation in the optical-UV regime to produce the necessary photoionization of the BLR. We also note that the optical polarization is not well correlated with \g-ray flux in \ca0454\ as in \thesource\ and \df1510, suggesting a different physical mechanism causes the line variability in \ca0454.

\section{Conclusions}\label{sec:samcon}
Over 5 years of observations, we find little statistically significant emission line variability in the 7 sources presented here. However, in 3 \g-ray flaring blazars, \ca0454, \thesource\ and \df1510, significant emission line variability was detected. We test the optical polarization as a proxy for non-thermal jet contribution during periods of emission line variability in these sources. We find that in some cases the optical polarization increases with the \fermi\ \g-ray flux and emission line flares, but the correlation is poor, probably because the details of the jet structure affect the polarization signal. From the \g-ray flaring, we infer the presence of non-thermal photoionizing photons in the system that could interact with the broad line region and cause the observed emission line increases.

While we cannot conclusively determine whether there are lags between the emission line increases and the \g-ray flares, due to poor temporal sampling in the present dataset, we find that the most \g-ray active blazars have statistically significant line variability that is not seen in less \g-ray active sources. 

Higher cadence optical spectroscopy is needed to investigate better the correlation between the emission lines and the \g-ray fluxes. This requires near daily spectroscopy in both active and quiescent states to build up enough data to truly constrain the degree of correlated variability between the jet and emission line flux.

\acknowledgements
\textit{Acknowledgements.} We thank the referee for insightful comments and recommendations that improved the quality of this work. We also wish to thank Benoit Lott for the use of the adaptive binning code, Alan Marscher for fruitful discussions, as well as Tanguy Marchand and Connor Hoge for contributing to the group discussions of this work. \smarts\ observations of LAT-monitored blazars are supported by 
Yale University and \fermi\ GI grant NNX14AQ24G. J.C.I. receives support from the Chancellor's Faculty Fellowship (Syracuse University's NSF ADVANCE grant HRD-1008643). The Steward Observatory blazar monitoring project is supported by Fermi Guest Investigator grants NNX08AW56G, NNX09AU10G, and NNX12AO93G.

\appendix
\section{\smarts\ Optical and Infrared Finding Charts}
\label{app:A}
The \smarts\ optical and infrared finding charts for \ba0402, \ca0454, \e2052, and \f2142\ are presented here with the BVRJK magnitudes used to calibrate the comparison stars.  \smarts\ OIR finding charts for \df1510\ and \a0208\ have been previously published in \citet{Bonning12}. All \smarts\ OIR finding charts can be found on our website.

\begin{figure*}
\plottwo{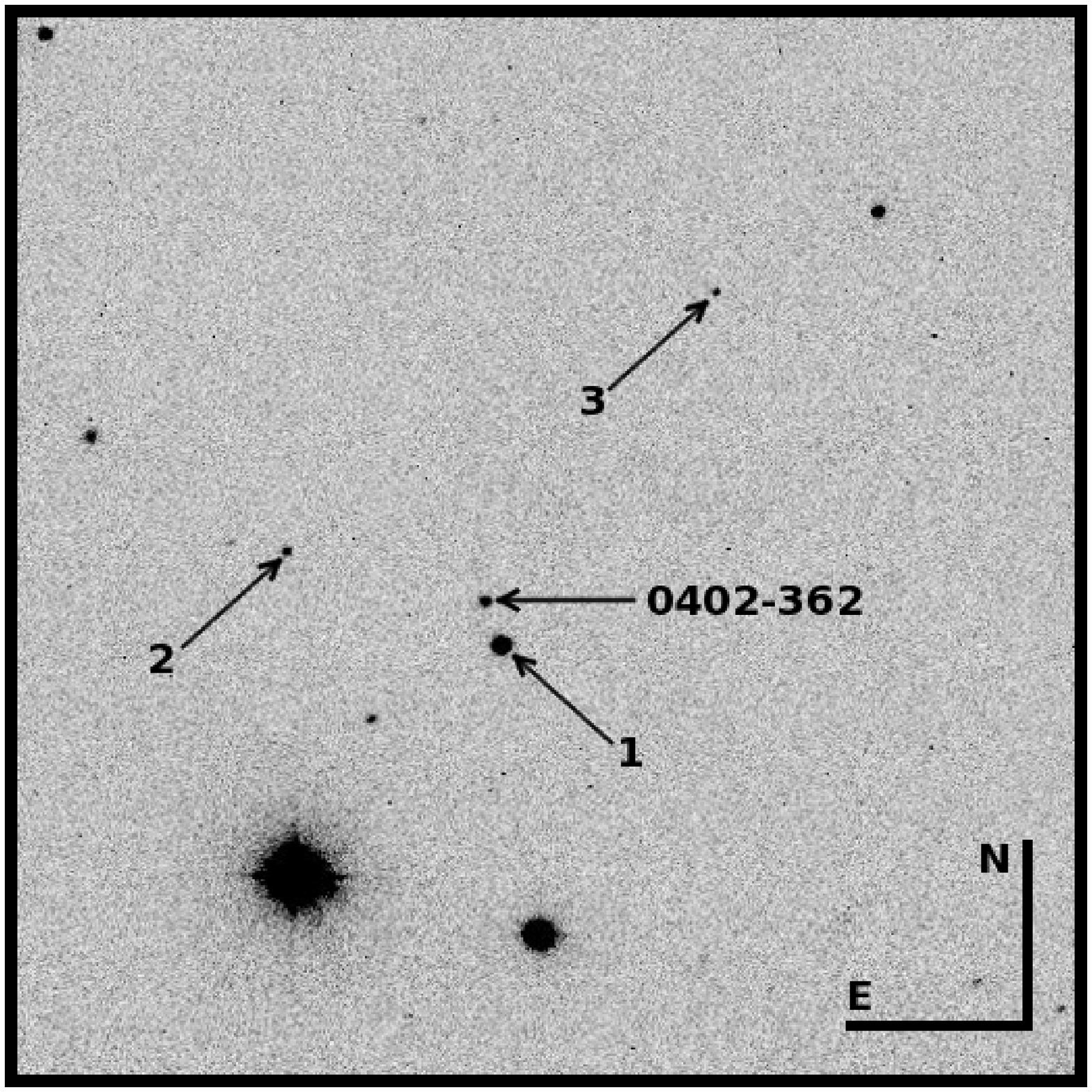}{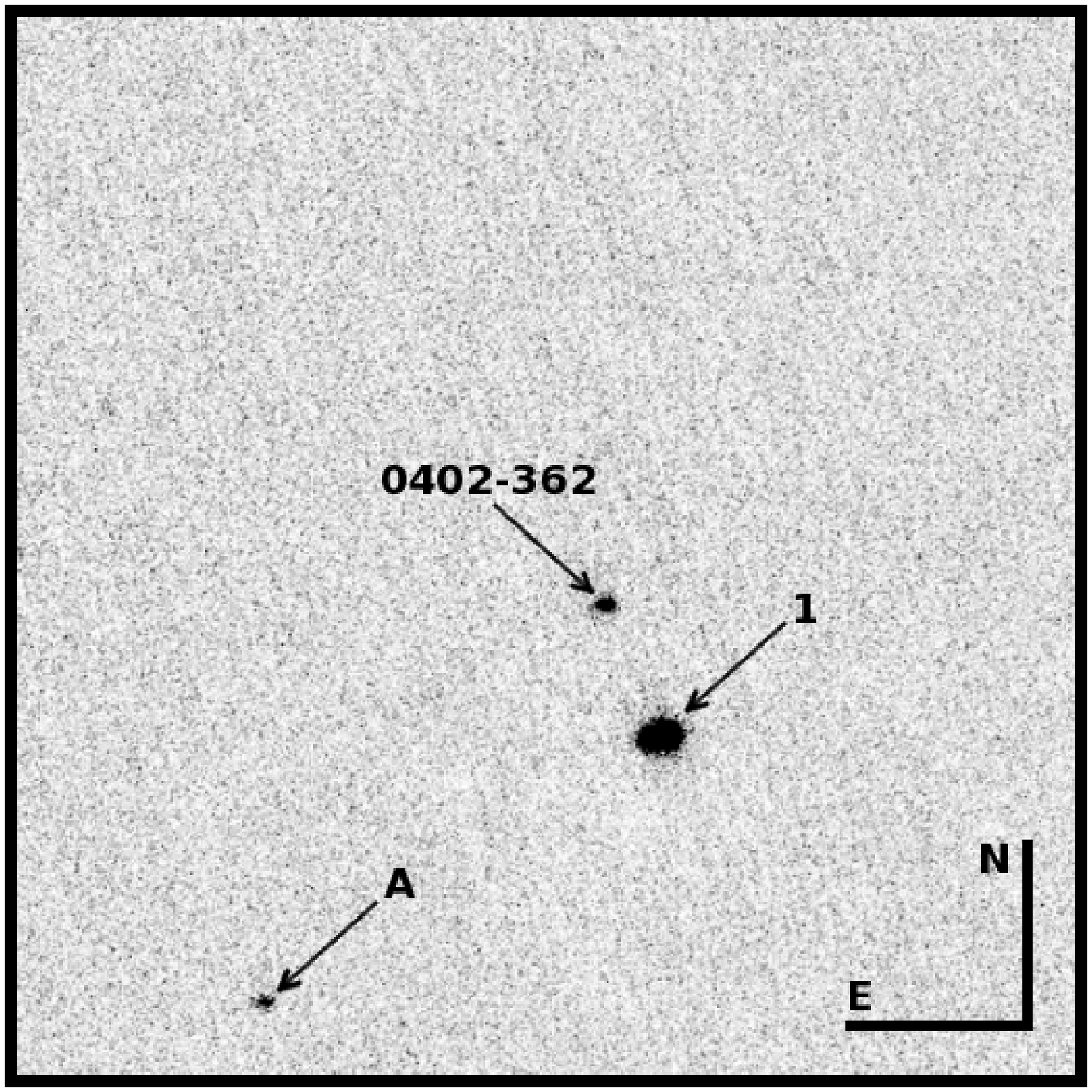}
\caption{The \smarts\ optical (left) and infrared (right) finding charts for \ba0402, \ca0454, \e2052, and \f2142.  The field of view is 5.12$^\prime$ $\times$ 5.12$^\prime$ for the optical finding charts and 1.45$^\prime$ $\times$ 1.45$^\prime$ for the infrared. Comparison star magnitudes and 1$\sigma$ uncertainties are given in Table~\ref{tab:samcomps}; optical comparison stars are labeled with numbers and infrared comparison stars are labeled with letters. In cases where optical and infrared comparison stars are the same, numbers are used to identify the star in both images. \label{fig:a040}}
\end{figure*}

\begin{figure*}
\plottwo{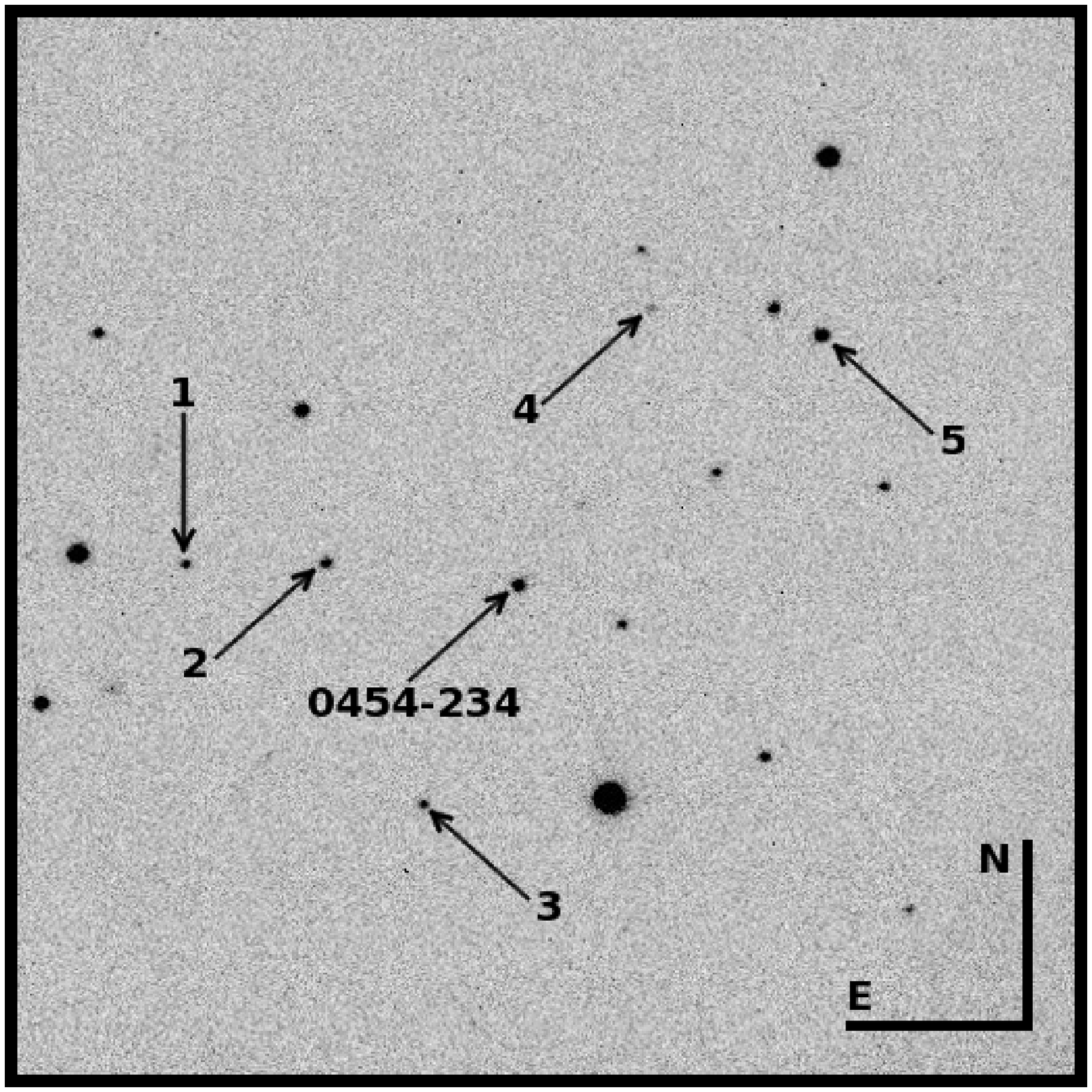}{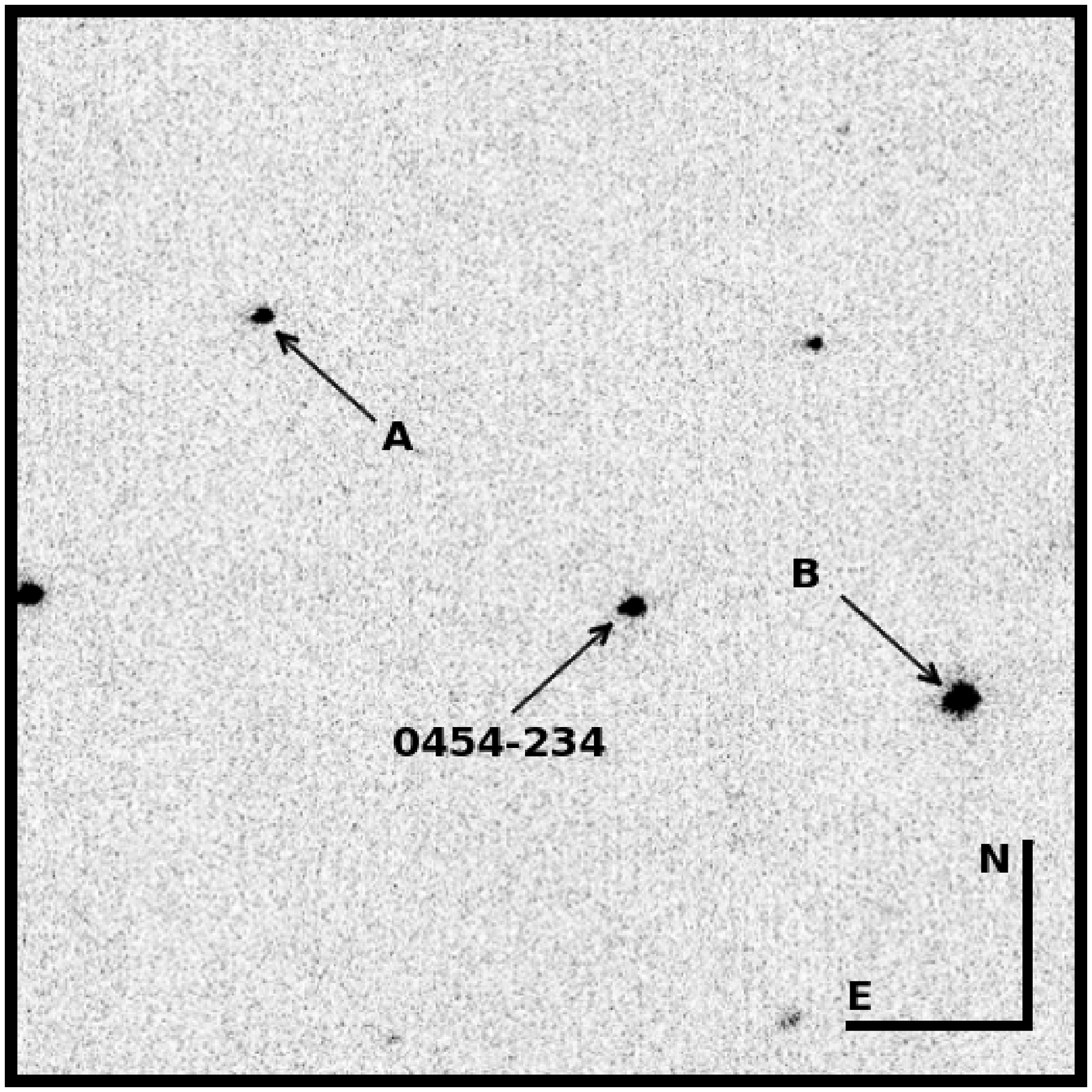}
\caption{Optical and infrared finding charts for \ca0454, labels as in Figure~\ref{fig:a040}.}
\end{figure*}

\begin{figure*}
\plottwo{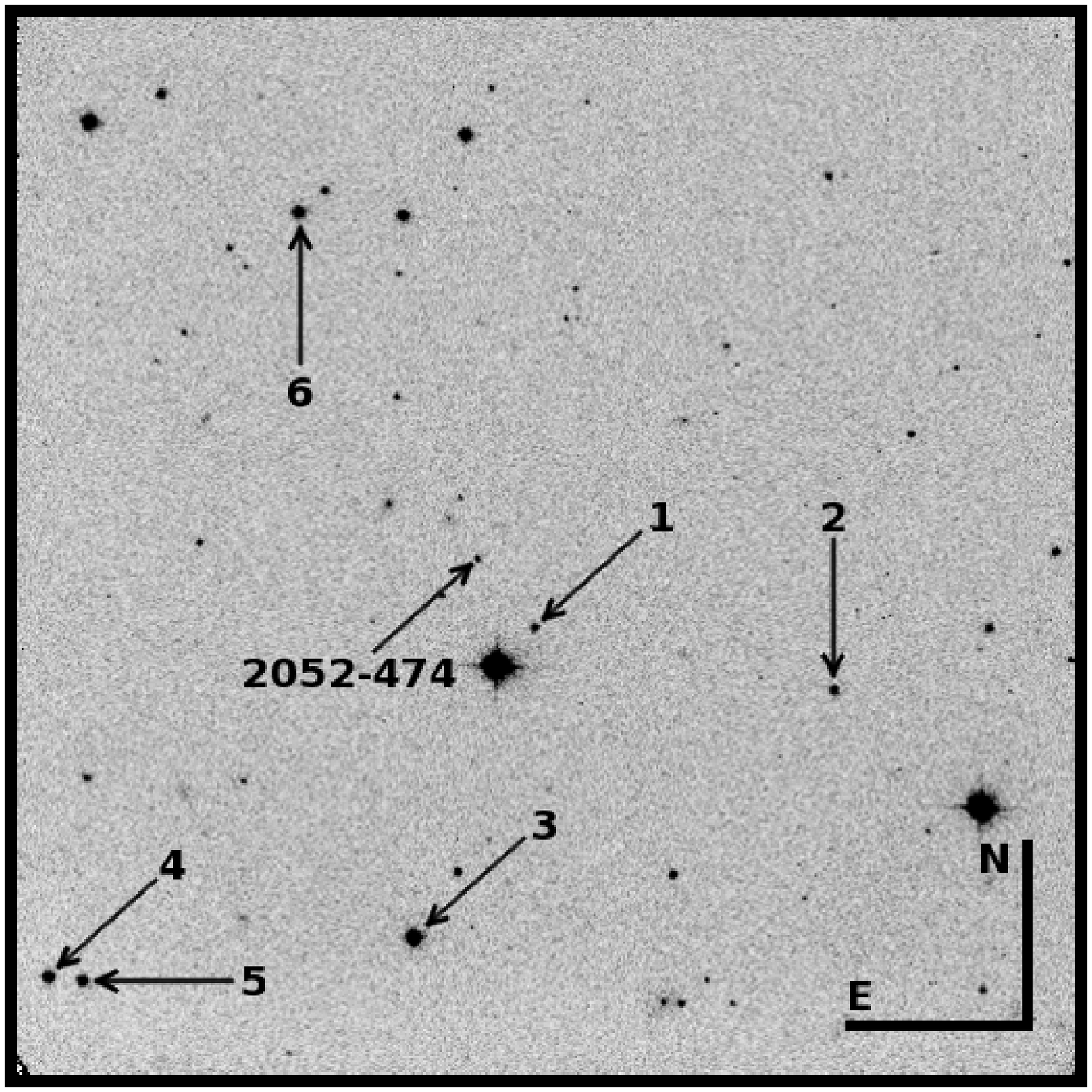}{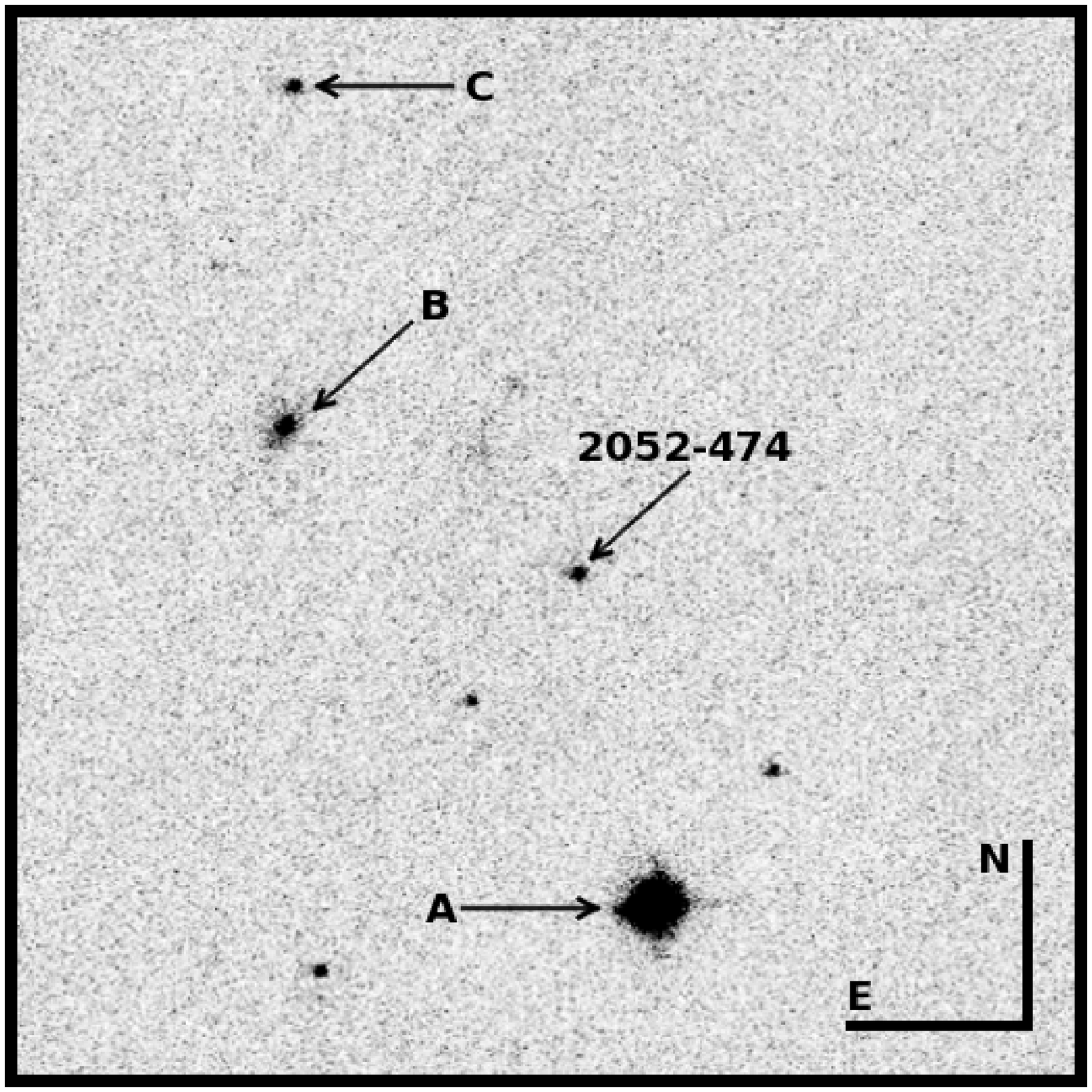}
\caption{Optical and infrared finding charts for PKS~2052-474, labels as in Figure~\ref{fig:a040}.}
\end{figure*}

\begin{figure*}
\plottwo{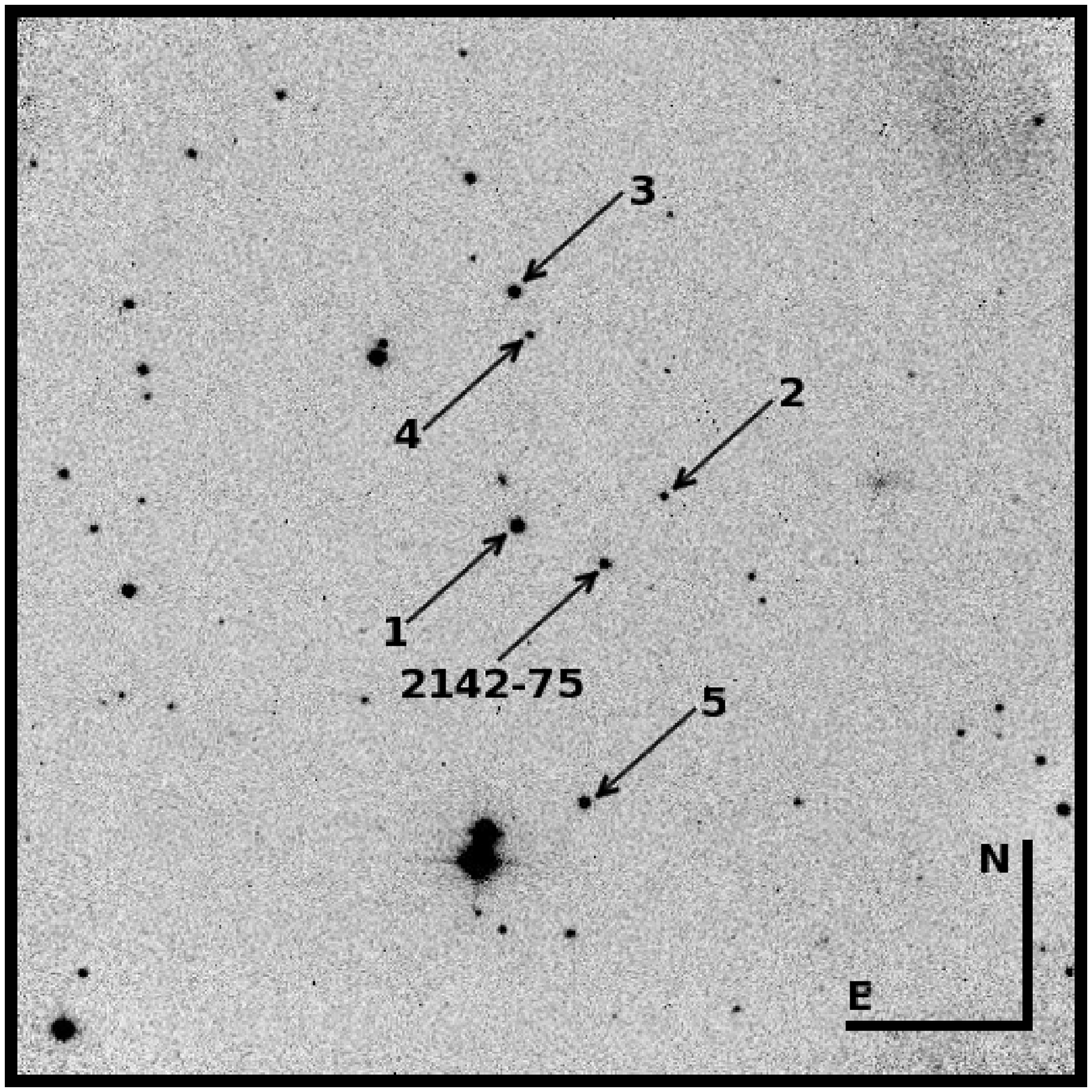}{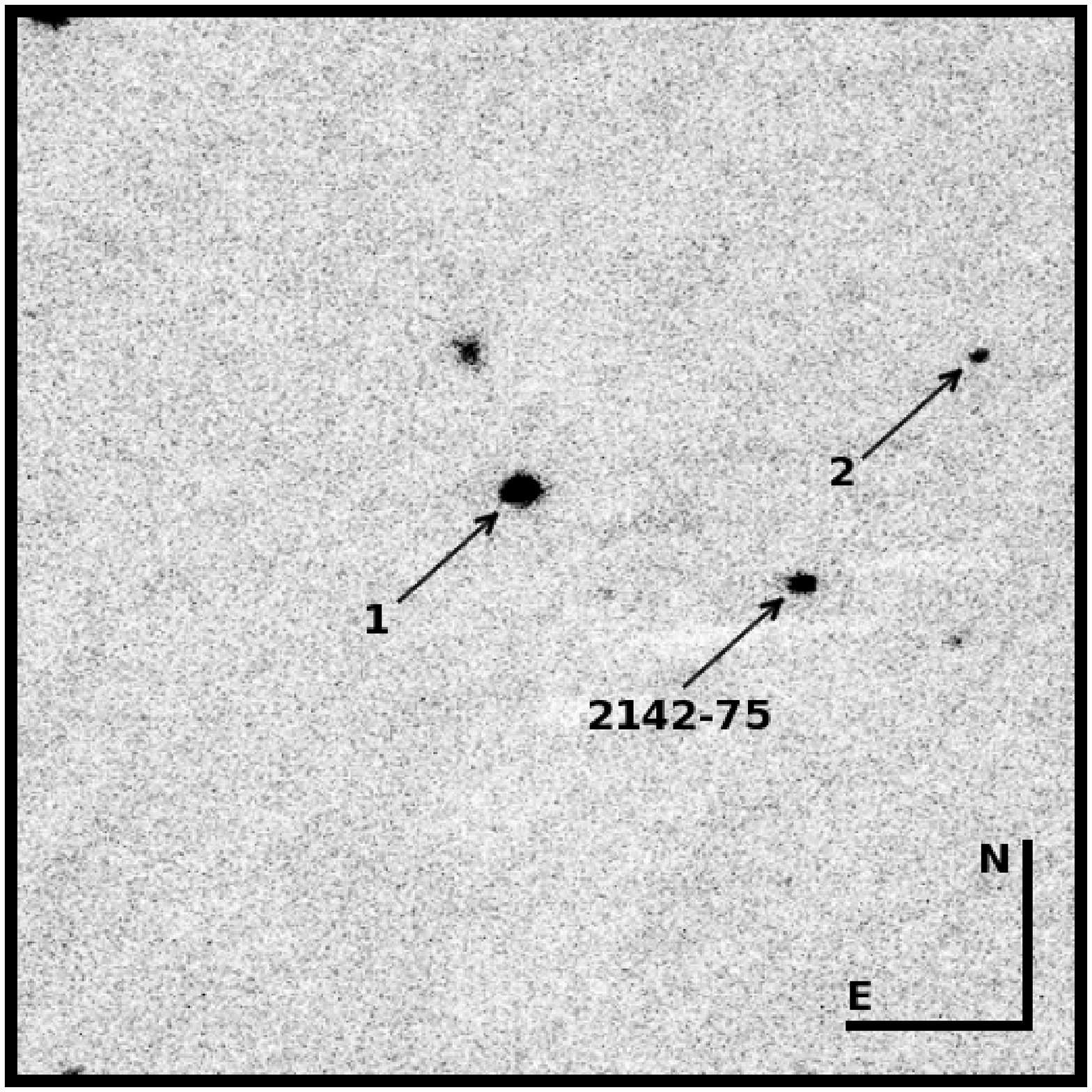}
\caption{Optical and infrared finding charts for PKS~2142-474, labels as in Figure~\ref{fig:a040}.}
\end{figure*}

\begin{deluxetable*}{llllllll}

\tablecaption{Optical and Infrared Comparison Stars}
\tablehead{\colhead{Target} & \colhead{Star} & \colhead{B ($\sigma_B$)} & \colhead{V ($\sigma_V$)} & \colhead{R ($\sigma_R$)} & \colhead{J ($\sigma_J$)} & \colhead{K ($\sigma_K$)}}
\startdata
PKS 0402-362 & 1 & 15.38 (0.03) & 14.75 (0.03) & 14.34 (0.02) & 13.42 (0.03) & 13.02 (0.03) \\
 & 2 & 18.47 (0.06) & 17.14 (0.04) & 16.18 (0.02) & - & - \\
 & 3 & 19.17 (0.09) & 17.81 (0.03) & 16.81 (0.03) & - & - \\
& A & - & - & - & 16.82 (0.14) & 15.82 (...)  \\\hline
PKS 0454-234 & 1 & 18.04 (0.04) & 17.02 (0.03) & 16.32 (0.03) & - & - \\
 & 2 & 17.53 (0.03) & 16.55 (0.02) & 15.86 (0.02) & - & - \\
 & 3 & 18.12 (0.04) & 17.05 (0.03) & 16.32 (0.02) & - &- \\
 & 4 & 19.22 (0.06) & 17.83 (0.03) & 16.82 (0.03) & - & -\\
 & 5 & 16.65 (0.02) & 16.00 (0.02) & 15.62 (0.02) & - & -\\
 & A & - & - & - & 15.60 (0.06) & 14.86 (0.12) \\
 & B & - & - & - & 13.45 (0.02) & 12.65 (0.03) \\\hline
PKS 2052-474 & 1 & 18.29 (0.05) & 17.87 (0.10) & 17.52 (0.05) & - & - \\
 & 2 & 17.51 (0.02) & 16.94 (0.01) & 16.59 (0.03) & - & - \\
 & 3 & 15.28 (0.02) & 14.67 (0.01) & 14.29 (0.02) & - & - \\
 & 4 & 16.43 (0.02) & 15.82 (0.01) & 15.46 (0.03) & - & - \\
 & 5 & 16.99 (0.02) & 16.23 (0.02) & 15.80 (0.03) & - & - \\
 & 6 & 16.25 (0.02) & 15.56 (0.01) & 15.15 (0.01) & - & - \\
 & A & - & - & - & 11.44 (0.02) & 11.24 (0.02) \\
 & B & - & - & - & 15.78 (0.11) & 14.43 (0.10) \\
 & C & - & - & - & 16.43 (0.11) & 16.67 (...)  \\\hline
PKS 2142-75 & 1 & 17.25 (0.04) & 16.12 (0.02) & 15.41 (0.02) & 13.90 (0.03) & 12.96 (0.04) \\
 & 2 & 19.11 (0.08) & 18.28 (0.03) & 17.75 (0.04) & 16.51 (0.14) & 15.84 (...) \\
 & 3 & 17.23 (0.04) & 16.51 (0.02) & 16.04 (0.03) & - & - \\
 & 4 & 19.23 (0.06) & 18.39 (0.04) & 17.90 (0.04) & - & - \\
 & 5 & 17.98 (0.04) & 16.93 (0.03) & 16.18 (0.03) & - & - 
\enddata
\tablecomments{The reported uncertainties are 1$\sigma$. Optical comparison star magnitudes are listed by number, and infrared comparison star magnitudes are listed by letter. When the same comparison star is used in the optical and infrared, the requisite data is labeled by number for both bands. Infrared uncertainties in 2MASS that were not available via the catalog are denoted by `...'. These data are calibrated to the 23 June 2014 optical and infrared photometry. \label{tab:samcomps}}
\end{deluxetable*}

\bibliography{refs2012}
\end{document}